\RequirePackage{amsmath}
\documentclass[10pt]{iopart}
\usepackage{graphicx}
\usepackage{epsfig}
\usepackage{amsfonts}
\usepackage{amsmath}
\usepackage{amssymb}
\usepackage{subfigure}
\usepackage{color}
\usepackage{multirow}
\usepackage{array}

\newcommand{\ct}{\cite}
\newcommand{\be}{\begin{equation}}
\newcommand{\ee}{\end{equation}}
\newcommand{\ba}{\begin{eqnarray}}
\newcommand{\ea}{\end{eqnarray}}

\newcommand{\non}{\nonumber}

\newcommand{\de}{\delta}

\begin{document}
\title{Stability and pre-thermalization in chains of classical kicked rotors}
\author{Atanu Rajak$^{1}$, Roberta Citro$^{2}$, and Emanuele G. Dalla Torre$^{1}$}
\ead{raj.atanu009@gmail.com} 
\address{  
 $^{1}$Department of Physics and QUEST, Bar-Ilan University, Ramat-Gan 52900, Israel\\
 $^{2}$Dipartimento  di  Fisica  ”E.R.  Caianiello”,  Universit$\grave{a}$  degli  Studi  di  Salerno  and  Spin-CNR  Unit$\grave{a}$,
Via  Giovanni  Paolo  II,  132,  I-84084  Fisciano  (Sa),  Italy
 }

\begin{abstract}

Periodic drives are a common tool to control physical systems, but have a limited applicability because 
time-dependent drives generically lead to heating. How to prevent the heating is a fundamental question 
with important practical implications. We address this question by analyzing a chain of coupled kicked rotors, 
and find two situations in which the heating rate can be arbitrarily small: (i) linear stability, for initial 
conditions close to a fixed point, and (ii) marginal localization, for drives with large frequencies 
and small amplitudes. In both cases, we find that the dynamics shows universal scaling laws that allow us to 
distinguish localized, diffusive, and sub-diffusive regimes. The marginally localized phase has common traits 
with recently discovered pre-thermalized phases of many-body quantum-Hamiltonian systems, but does 
not require quantum coherence.

\end{abstract}
\maketitle

\section{Introduction}
\label{intro}
Periodically driven many-body systems are currently being studied intensively, both in theory 
and in experiments. One common research direction, known as Floquet engineering, uses  periodic 
drives to generate tunable effective couplings. For example, periodic drives can create negative 
kinetic energies, leading to dynamical localization and kinetic frustration in experiments on ultracold atoms 
(see Ref.~\cite{eckardt17atomic} for a recent review). Similarly, periodically oscillating phonons 
are believed to be responsible for the light-enhanced coherence observed in superconductors
\cite{fausti11light,hu13optically,babadi2017theory,sentef2017light,kennes2016electronic,murakami2017nonequilibrium}. 
Periodic drives can additionally be used to create phases of matter that do not have any equilibrium counter-part, 
such as Floquet topological phases \cite{oka09photovoltaic,lindner11floquet,kitagawa11transport,rechtsman13photonic,rudner13anomalous,jotzu14experimental,fleury16floquet} and
Floquet time crystals \cite{khemani16phase,else16floquet,zhang17observation,moessner2017equilibration,russomanno2017spin,yao17discrete}.

The key difference between time-independent and periodically driven Hamiltonian systems is that in the latter, 
the energy is not conserved. As a consequence, Floquet systems are generically subject to heating, as their 
average energy increases with time. For many-body closed quantum systems, one can rely on an extension of 
the eigenstate thermalization hypothesis (see Ref.~\cite{mori2017thermalization} for an introduction), which is
valid at asymptotically long times. Because the energy is not conserved, periodically driven systems 
are expected to effectively thermalize 
at an infinite temperature \cite{d13many,d14long,lazarides14equilibrium,russomanno15asymptotic,russomanno15thermalization,seetharam15controlled}. 

Recent studies showed that in some cases, the heating can be reduced, or even canceled \cite{prosen98time,d13many,citro15dynamical,bhadra2018dynamics}.  
For example, it was found that the heating can be prevented if the system is integrable \cite{russomanno12periodic,russomanno2016kibble}, 
or ``many-body localized'' \cite{ponte15many,lazarides15fate,ponte15periodically,abanin16theory,agarwal17localization,dumitrescu17logarithmically}. 
Alternatively, the heating can be suppressed if the driving frequency is larger than the single-particle band-width 
\cite{choudhury14stability,bukov15prethermal,abanin15exponentially,citro15dynamical,goldman15periodically,chandran16interaction,mori2016rigorous,
abanin17rigorous,lellouch17parametric,lellouch17parametric1,machado17exponentially}. Under these conditions, the system can show 
interesting long-lived pre-thermal states~\cite{kuwahara16floquet,canovi16stroboscopic,weidinger17floquet,zeng17prethermal,else17prethermal,abanin17effective}. 
The common feature of all these studies is that they deal with {\it quantum} Hamiltonian systems.

In this article, we explore a {\it classical} periodically driven many-body system, by considering a chain 
of coupled kicked rotors. 
As we will see, this simple system shows many features that were predicted earlier for the more 
complex quantum case. Our model is a natural extension of the single kicked rotor, whose dynamics 
is described by the well-known ``Chirikov standard map''~\ct{chirikov79a}. This single-particle 
model offers a paradigmatic example of a transition between regular dynamics and chaos, or equivalently 
localization and diffusion. Earlier studies revealed that when many rotors are coupled together, the 
transition is washed out: at exponentially long times, 
many-body kicked rotors are generically diffusive~\cite{kaneko89diffusion,konishi90diffusion}.

By focusing on the finite-time dynamics, we find regions in the parameter space that are localized, 
diffusive, and super-diffusive. These three regimes are characterized by a different behavior of 
the average kinetic energy of the rotors. In the localized regime, this quantity does not grow with 
time, indicating that the system does not heat up. In contrast, in the diffusive regime, the system 
absorbs energy at a constant rate. 
The transition between the localized and the diffusive regimes depends on the initial conditions. 
If the system is initialized close to a stable point, the transition corresponds to a sharp resonance, which 
does not drift with increasing time. If the system is instead initialized with large fluctuations, the critical 
coupling of the transition depends logarithmically on time.
When logarithmic corrections are taken into account, the diffusion parameters show a universal asymptotic scaling, 
as a function of the kick strength and frequency.

The paper is organized as follows: We first briefly review the physics of a single kicked rotor 
(Sec.~\ref{skr}) and present its many-body generalization (Sec.~\ref{mkr_cd}). Then, we describe 
the case where the rotors are prepared in vicinity of a stable fixed point, and discuss the system's 
linear stability and resonances, both analytically (Sec.~\ref{qe}) and numerically (Sec.~\ref{ls}). 
Next, we address generic initial conditions and study the regime of marginal localization (Sec.~\ref{ms}). 
Finally, we characterize the different regimes using a Fourier decomposition 
(Sec.~\ref{msr}), and draw our conclusions (Sec.~\ref{conclusion}).

\section{Review of a single kicked rotor}
\label{skr}
The Hamilton of a single kicked rotor is given by
\be
H=\frac{p^2}{2}-\kappa\cos{(\phi)}\Omega(t), \text{with} \hspace{0.2cm} \Omega(t)=\sum_{n=-\infty}^{+\infty}\de(t-n\tau).
\label{ham_skr}
\ee
Here, the position of the rotor is described by the angle $\phi$, and the 
associated momentum by $p$. The parameters $\kappa$ and $\tau$ are the strength and the period of the kicks, respectively.
For $\kappa>0$, the static model ($\Omega(t)\to\Omega_0>0$) has a stable fixed point at $p=\phi=0$, 
and a unstable one at $p=0,\phi=\pi$. Note that this model can be mapped to the $\kappa<0$ model under 
the transformation $\phi \to \phi + \pi$, which flips the role of the stable and unstable fixed points. 
Thus, without the loss of generality, we focus here on the case $\kappa>0$ only.

We now briefly review the linear analysis of Eq.~(\ref{ham_skr}) for the sake of completeness, and due to its relevance 
to the many-body case. We expand $\cos \phi$ around 
$\phi=0$ and apply the quadratic approximation $\cos \phi\approx 1-\phi^2/2$  
in Eq.~(\ref{ham_skr}). The classical equations of motion are then equivalent to a kicked harmonic oscillator\cite{weigert02quantum}, where
\be
\frac{d}{dt}\begin{pmatrix} \phi \\ p \end{pmatrix}=\begin{pmatrix} 0 & 1\\ -\kappa\Omega(t) & 0 \end{pmatrix}\begin{pmatrix} \phi \\ p \end{pmatrix}.
\label{eq_motion}
\ee

We consider the evolution of the system over one time period, from $t=-\varepsilon$ to $t=\tau-\varepsilon$ 
with $\varepsilon\ll \tau$.
The solution of Eq.~(\ref{eq_motion}) is given by
\be
\begin{pmatrix} \phi(\tau-\varepsilon) \\ p(\tau-\varepsilon) \end{pmatrix}=\varUpsilon\exp\Big[\int_{-\varepsilon}^{\tau-\varepsilon}dtM(t)\Big]
\begin{pmatrix} \phi(-\varepsilon) \\ p(-\varepsilon) \end{pmatrix}
\ee
where $\varUpsilon$ denotes time ordering and $M(t)$ is $2\times2$ matrix of Eq.~(\ref{eq_motion}). 
During the first part of the period, for $-\varepsilon<t<+\varepsilon$, the rotor is kicked and the time evolution is determined by the matrix
\be
M_k=\lim_{\varepsilon\rightarrow0}\exp\Big[\int_{-\varepsilon}^{+\varepsilon}dtM(t)\Big]=\exp\begin{pmatrix} 0 & 0\\ -\kappa & 0 \end{pmatrix}
=\begin{pmatrix} 1 & 0\\ -\kappa & 1 \end{pmatrix}.
\ee
In the second part of the time period, $\varepsilon<t<\tau-\varepsilon$, the particle experiences a free motion, described by the matrix
(with $\varepsilon\rightarrow0$)
\be
M_0=\begin{pmatrix} 1 & \tau\\ 0 & 1 \end{pmatrix}.
\ee
As a result, the phase space evolution of the kicked rotor over one time period is given by the matrix
\be
M=M_0M_k=\begin{pmatrix} 1-\kappa \tau & \tau\\ -\kappa & 1 \end{pmatrix}.
\label{matrixM}
\ee

The transition between the stable and unstable dynamics of the oscillator can be inferred from the 
eigenvalues of the matrix $M$, Eq. (6), which are given by
\ba
\lambda_{\pm}
&=&\Big(1-\frac{\kappa \tau}{2}\Big)\pm\sqrt{\Big(1-\frac{\kappa \tau}{2}\Big)^2-1}.
\ea
We find that the properties of the eigenvalues are determined by the unitless parameter $K=\kappa \tau$: for $K<4$
the eigenvalues are a complex conjugate pair $\lambda_{\pm}=\exp(\pm i \theta)$ with $\cos \theta=1-K/2$, while for $K>4$ they are a real 
reciprocal pair $\lambda_{\pm}=\exp(\pm  \beta)$ with $\cosh \beta=1-K/2$. 
Thus, the point $K=4$ separates the stable ($K<4$) and unstable ($K>4$) regions, and can be associated with 
a parametric resonance of the system \footnote{The distinction between the stable and unstable regime is due 
to the fact that the matrix $M$ is real and has determinant ${\rm det} M = 1$. These conditions lead to two 
distinct options only: either both eigenvalues are real and have reciprocal values ($\lambda$ and $1/\lambda$), 
or they are pure phases ($e^{i\theta}$ and $e^{-i\theta}$). The transition between these two situations occurs 
at an exceptional point, where the two eigenvalues coalesce (i.e. are both equal to $1$, or to $-1$). 
Physically, this point describes the transition between the stable and unstable regimes of the oscillator.}.
Note that the unitless parameter $K=\kappa\tau$ sets the ratio between the 
driving amplitude and the driving frequency: as expected, the system 
is stable for weak and fast drives, and unstable for strong and slow drives.

The model (\ref{ham_skr}) is furthermore known to undergo a transition between a
localized and a diffusive regime at $K= K_c \approx 0.9716$~\ct{greene79method}. 
This phenomenon can be explained analyzing the evolution of the phase space as a function of $K$. 
For $K=0$, the system is integrable and $p$ is a conserved quantity. In this case, the orbits are horizontal lines 
that slice the phase space in distinct regions. According to the Kolmagorov-Arnold-Moser (KAM) theorem~\cite{arnold68}, for 
small values of $K$, the system's dynamics is governed by stable invariant submanifolds (KAM tori). 
These KAM tori divide the phase space into many sections, 
each of which may carry stochastic orbits. Therefore the momentum is bounded between these invariant 
tori and the kinetic energy does not grow with time (localized regime). As the value of $K$ is increased, 
the number of KAM surfaces decreases and, at some critical 
value $K=K_c$, the last KAM tori disappears.
For $K<K_c$ the momentum of the pendulum 
is localized between invariant these KAM tori, independently of the initial conditions, while for $K>K_c$ the system 
can show a diffusive behavior (in momentum space) for some initial conditions. Note that this transition occurs 
in the region where the point $\phi=0$ is linearly stable. 
The distinction between linear stability and localization will be a key ingredient for our analysis of the many-body case.

The Hamiltonian in Eq.~(\ref{ham_skr}) can be quantized, by assuming $p$ and $\phi$ to be quantum operators 
satisfying canonical conjugation relations. For this system, the quantum evolution operator over one period is 
the product of free evolution and instantaneous kick operators,
\begin{equation}
U=\exp(-ip^2\tau/2)\exp(i\kappa\cos\phi)
\end{equation}
which is called Floquet operator. The stroboscopic evolution of the system is completely determined 
by the eigenstates of the Floquet operator. Using the eigenvalue equation of the Floquet operator, it has been 
shown before that a quantum kicked rotor can be mapped to a quantum particle moving in a one-dimensional static disordered potential
~\ct{fishman82chaos,fishman89scaling}.

Because of Anderson localization~\cite{anderson58absence}, all the eigenfunctions are localized in momentum space 
and the kicked rotor cannot diffuse. Thus, the quantization of Eq. (1) leads to a model that is always localized, 
irrespective of the value of K. This effect is termed ``dynamical localization'' and is due to the quantum coherence 
between the different trajectories in momentum space.


Recently, the effect 
of quantum interferences in the dynamics of $N$ interacting kicked rotors was investigated by mapping this model to a single quantum particle 
in a $N$-dimensional disordered lattice~\ct{notarnicola17from}. For $N>2$, this model shows a transition between Anderson localized 
and delocalized phases, but the critical coupling tends to zero as $N$ increases. Thus, in the thermodynamic 
limit of $N\to \infty$, the effect of quantum coherence is negligible and one would expect the model to be always diffusive \ct{seetharam17absence}, 
as in the classical case \ct{kaneko89diffusion,konishi90diffusion}.

\section{Classical many-body kicked rotor}
\label{mkr_cd}
We consider a chain of $N$ coupled kicked rotors, defined by the Hamiltonian
\be
H~=~\sum_{j=1}^N\Big[\frac{p_j^2}{2}-\kappa\cos(\phi_j-\phi_{j+1})\sum_{n=-\infty}^{+\infty}\de(t-n\tau)\Big].
\label{ham_mkr}
\ee
This model is a natural extension of the kicked rotor (\ref{ham_skr}), where the coupling between the rotors is induced by the kick~
\footnote{The model (\ref{ham_mkr}) is very convenient for numerical 
analysis, compared to a model with sinusoidal periodic driving. In this latter case the time evolution needs 
to be discretized in infinitesimal steps, preventing the calculation of physical observables at very long times. 
In contrast, for the kicked rotor model we can go up to very long times without facing 
such problem, thanks to the discrete nature of its time steps. 
We believe that the physical properties of the system will not change depending on 
the particular type of driving protocol.}.
Without the loss of generality, 
we assume that $\kappa>0$: Models with positive and negative $\kappa$'s can be mapped to each other 
through the transformation $\phi_j \to \phi_j + j \pi$ for all $j=1,2,3,\cdots$. 
Using the classical Hamilton's equations of motion for the Hamiltonian in Eq.~(\ref{ham_mkr}), 
one obtains a relation between the phase space coordinates just before the kick $n$ and the same coordinates after one time period:
\begin{align}
p_j(n+1)=&~p_j(n)-\kappa\big[\sin(\phi_j(n)-\phi_{j+1}(n)) \non \\
&~~~~~~~~~~-\sin(\phi_{j-1}(n)-\phi_j(n))\big],\non\\
\phi_j(n+1)=&~\phi_j(n)+p_j(n+1)\tau. \label{cd_equs1} 
\end{align}
As mentioned in Sec.~\ref{skr}, one can find an unitless parameter $K=\kappa\tau$ such that 
the dynamics depends on $K$ only.

A possible realization of the interacting kicked rotor model is based on the use of superconducting grains or molecules. 
In the first case, the Hamiltonian, Eq.~(\ref{ham_mkr}), could be implemented using a chain of voltage-biased superconducting grains, 
in which the “kick” term is obtained by suddenly switching on a strong Josephson coupling for short times~\cite{keser16dynamical}. 
In the case of molecules, the Hamiltonian (\ref{ham_mkr}) could be 
implemented by periodic trains of laser pulses~\cite{cryan09field,zhdanovich12quantum,zahedpour14quantum}.
Alternatively, interacting kicked rotors can be realized using time-modulated potentials with non-commensurate 
wavelength~\cite{gadway13evidence}.

The model (\ref{ham_mkr}) has been studied before in the context of chaos in almost-integrable models (The model is trivially 
integrable at $K=0$)~\ct{chirikov97arnold,falcioni91ergodic,mulansky11strong}. These earlier studies focused on two aspects 
of the problem: (i) the distribution of the 
Lyapunov exponents $\lambda$; (ii) the scaling of the diffusion coefficient $D$ and of the largest $\lambda_{\rm max}$, with $K$. 
It was found that $\lambda_{\rm max}\sim\sqrt{K}$ and $D \sim K^\alpha$ with $\alpha \approx 6.5$~\ct{mulansky11strong}. 
This situation was named ``fast Arnold diffusion'', in contrast to the common Arnold diffusions, where $D\sim \exp(a/K^b)$ 
with $a,b>0$~\ct{chirikov79universal}. These studies focused on exponentially long times, where the system shows a diffusive behavior. 
In contrast, we here consider intermediate time scales, where the diffusion coefficient has not reached its long-time limit yet, 
and the dynamics still depends on the initial conditions. 

In this context, our work bears some analogies with the celebrated Fermi-Pasta-Ulam-Tsingou (FPUT) model~\cite{fermi65,berman05fermi}. 
The FPUT problem originated from one of the first numerical simulations of a physical system on a digital computer 
and involved the dynamics of a classical many-body system under a non-linear time-independent Hamiltonian. In contrast 
to the original expectations, the numerical solution performed by FPUT showed that the system was not ergodic, and did not 
lead to an equidistribution of the energy among the modes. This effect can be understood within the framework of Arnold 
diffusion. Because the non-linearities considered by FPUT were relatively small, the time required to observe diffusion 
were exponentially large and could not been observed within the computational resources available at that time.


In our numerical calculations, we consider an ensemble of initial conditions in which the initial momenta are zero, $p_j(0)=0$, 
and the initial positions are given by identical independent variable with a Gaussian distribution 
$f(\phi_j)=(\frac{1}{2\pi\sigma^2})^{\frac{1}{2}}\exp(-\frac{(\phi_j-j\phi_0)^2}{2\sigma^2})$, with mean 
$\langle\phi_j\rangle=j\phi_0$ and standard deviation $\sigma$. 
The parameter $\sigma$ sets the initial fluctuation of the system 
and will be an important tuning parameter of our theory.
In this paper, we will focus on two specific cases of $\phi_0$, the case $\phi_0=0$, where the system is initialized 
close to the classical ground state of the system ($\phi_j=p_j=0$), and the case $\phi_0=\pi$, where the nearest neighbor 
oscillators are initialized in opposite phases.

We characterize the dynamical phases of the model through the diffusion coefficient
\be
D=\frac{\langle p^2(t_f)\rangle-\langle p^2(t_i)\rangle}{t_f-t_i}.
\label{dc_def}
\ee
Here, $t_i$ and $t_f$ are two different stroboscopic times ($t=n\tau$, where $n$ is an integer), and 
$\langle p^2\rangle=\langle\frac{1}{N}\sum_{j=1}^{N}p_j^2\rangle$, where $\langle\cdots\rangle$ 
represents the averaging over initial conditions. 
All the averages in this work, are done with $1000$ different realizations of initial conditions 
(see also Appendix B for a case study of the error bars of our results).
Because the average kinetic energy of the model is $\frac12\langle p^2\rangle$, the coefficient $D$ measures the average heating rate of the system. 
This coefficient is expected to converge to a finite value in a diffusive phase, while it tends to zero in a localized phase. 
To distinguish between diffusive and super-diffusive phases, we further calculate
the exponent $\alpha$ by fitting the numerically computed $\langle p^2(t)\rangle$ with a generic power-law function $A t^{\alpha}$.
The parameter $\alpha$ allows us to identify phases that are diffusive ($\alpha=1$), subdiffusive ($\alpha<1$), and superdiffusive ($\alpha>1$).

Our main aim is to determine the stability diagram of the model (\ref{ham_mkr}). 
To achieve this goal we numerically determine the 
diffusion parameters $D$ and $\alpha$ 
as a function of $K$, for different choices of $\phi_0$ and $\sigma$. These quantities
are computed for different time ranges, where the initial time is fixed at 
$t_i=1000$ and the final time is varied up to $t_f=6.5\times 10^4$. The choice of $t_i$ is dictated by 
the demand that the energy redistributes between the position ($\phi_j$) and momentum ($p_j$) degrees of freedom 
(also see discussion in~\ref{appendixt}). 
Throughout the paper, we consider systems of $N=1000$ rotors. As shown in~\ref{apendixa}, our 
numerical results are not sensitive to finite size effects.

\section{Quadratic expansion and beyond}
\label{qe}
Before presenting our numerical results, let us first consider the linear stability of the model around $\phi_0=0$. 
Using the quadratic approximation
$\cos(\phi_j-\phi_{j+1})\approx 1-(\phi_j-\phi_{j+1})^2/2$. 
We can rewrite the Hamiltonian (\ref{ham_mkr}) as  a collection of decoupled kicked
harmonic oscillators 
\be
H=\frac{1}{2}\sum_q\Big[|P_q|^2+F(q)|\phi_q|^2\sum_{n=-\infty}^{+\infty}\de(t-n\tau)\big],
\label{quad_exp1}
\ee
where $q=2\pi k/N$, $k$ being an integer, is the wave number, and  
\be
F(q)=4\kappa\sin^2(\frac{q}{2}). 
\label{fq}
\ee
Here $P_q=1/\sqrt{N}\sum_{j=1}^Np_je^{-iqj}$ and $\phi_q=1/\sqrt{N}\sum_{j=1}^N\phi_je^{-iqj}$ 
are the Fourier transforms of $p_j$ and $\phi_j$ respectively. 
The Hamiltonian in Eq.~(\ref{quad_exp1}) is a set of decoupled harmonic oscillators and is therefore integrable 
for any value of $K$. This Hamiltonian can nevertheless describe the transition between the localized and delocalized 
region, as the appearance of dynamical instabilities, where the phase space orbits change from elliptic to hyperbolic.


To determine the dynamical stability of Eq.~(\ref{quad_exp1}), we proceed in analogy to Sec.~\ref{skr} and first derive the 
associated equations of motion. Note that since $p_j$ and $\phi_j$ are real functions, we have
$P_q^*=P_{-q}$ and $\phi_q^*=\phi_{-q}$. 
As a result, for each $q$, the classical equations of motion are given by 
\be
\frac{d}{dt}\begin{pmatrix} \phi_q \\ \phi_{-q} \\ P_q \\ P_{-q} \end{pmatrix}
=\begin{pmatrix} 0 & 0 & 0 & 1\\  0 & 0 & 1 & 0 \\ 0 & -F(q)\Omega(t) & 0 & 0 \\ -F(q)\Omega(t) & 0 & 0 & 0 \end{pmatrix}
\begin{pmatrix} \phi_q \\ \phi_{-q} \\ P_q \\ P_{-q} \end{pmatrix}.
\label{eq_motion_mkr}
\ee
Following a few steps similar to 
Sec.~\ref{skr}, we obtain the matrix $M_q$ for each $q$, which provides the time evolution of the system over one 
full period, and is given by \footnote{Note that the matrix $M_q$ is real and of unit determinant 
as in the case of the single rotor, (Eq.~(\ref{eq_motion})), although its size is $4\times 4$, rather than $2\times 2$.}
\be
M_q=\begin{pmatrix} 1-F(q)\tau & 0 & 0 & \tau\\  0 & 1-F(q)\tau & \tau & 0 \\ 0 & -F(q) & 1 & 0 \\ -F(q) & 0 & 0 & 1 \end{pmatrix}.
\ee

Diagonalizing the matrix $M_q$, we get two degenerate eigenvalues for each $q$:
\be
\lambda_{\pm}(q)=\Big(1-\frac{F(q)\tau}{2}\Big)\pm\sqrt{\Big(1-\frac{F(q)\tau}{2}\Big)^2-1}.
\label{ev2_mkr}
\ee
From Eq.~(\ref{ev2_mkr}), the condition of the transition between stable and unstable regions 
is given by $F(q)\tau=4$, or equivalently $K \sin^2(q/2) = 1$, for each $q$. 
As a consequence, the system is stable for $K<1$: in this regime, all the
eigenvalues are pure phases 
$\lambda_{\pm}(q)=e^{\pm i\theta(q)}$ with $\cos \theta(q)=1-F(q)\tau/2$.
For later reference, we rewrite the eigenvalues as $\lambda_{\pm}(q)=e^{\pm i\mu(q)\tau}$, 
where $\mu(q)=\theta(q)/\tau$ is the frequency of $q$-mode.

Although the system is linearly stable for $K<1$, higher-order effects can lead to non-linear resonances. 
To study these effects, let us consider the 4$^{\rm th}$ order term in the Taylor expansion of Eq.~(\ref{ham_mkr}) around $\phi_0=0$:
\begin{equation}
H^4 = -\frac{\kappa}{24}\sum_{j=1}^N(\phi_j-\phi_{j+1})^4\sum_{m=1}^\infty \cos\left(\frac{2\pi m}{\tau} t\right),\label{eq:V}
\end{equation}
where we rewrote the sum of delta functions (the kicks) as a Fourier series, and neglected the constant $m=0$ term. The ``unperturbed'' 
Hamiltonian is given by Eq.~(\ref{quad_exp1}) and describes a set of decoupled kicked harmonic oscillators. In this case, perturbation 
theory states that the system will absorb energy only if one of the driving frequencies equals the sum of the 
frequencies of 4 oscillators, or
\begin{equation}
\frac{2\pi m}{\tau} = \sum_{i=1}^4\mu(q_i)\;,\label{eq:sum}
\end{equation}
where $\mu(q)$ is the frequency of a $q$ mode as defined above.
Because $\mu(q)\le\mu(\pi)$, the right-hand side of Eq.~(\ref{eq:sum}) 
is smaller or equal than $4\mu(\pi)=4\cos^{-1}(1-2K)/\tau$. Thus, the resonance condition (\ref{eq:sum}) can be satisfied only if 
\begin{equation}
2\pi m\le4\cos^{-1}(1-2K).
\end{equation}
This condition can be satisfied only for $m=1$, and predicts that the system is unstable to quartic couplings for $K\ge 0.5$.
If we consider higher-order terms in the Taylor expansion of the cosine, we can obtain resonances at lower values of $K$: In general, 
the $k^{\rm th}$-order term $(\sim\phi^k)$ gives rise to a series of resonances\footnote{As a sanity check of our approach one can verify 
that second order terms $\sim (\phi_i-\phi_{i+1})^2$ give rise to a resonance at $K=m$, in agreement with the linear stability analysis.} at $K=1/2(1-\cos(2m\pi/k))$.



\begin{figure}[t]
\centering
\includegraphics[trim={1.4cm 0.5cm 2cm 0.7cm},clip,scale=0.46]{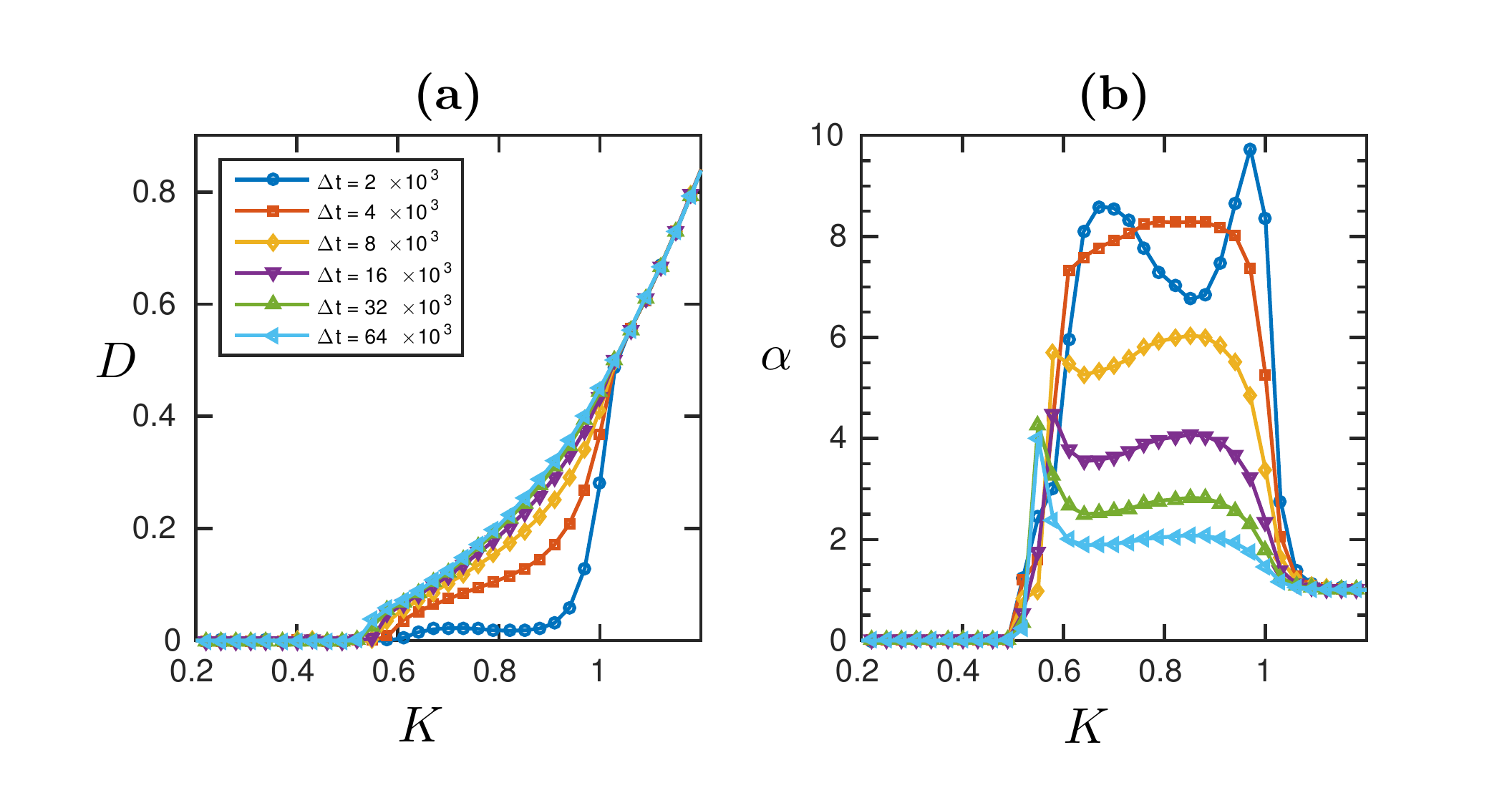}
\includegraphics[trim={1.4cm 0.5cm 2cm 0.7cm},clip,scale=0.46]{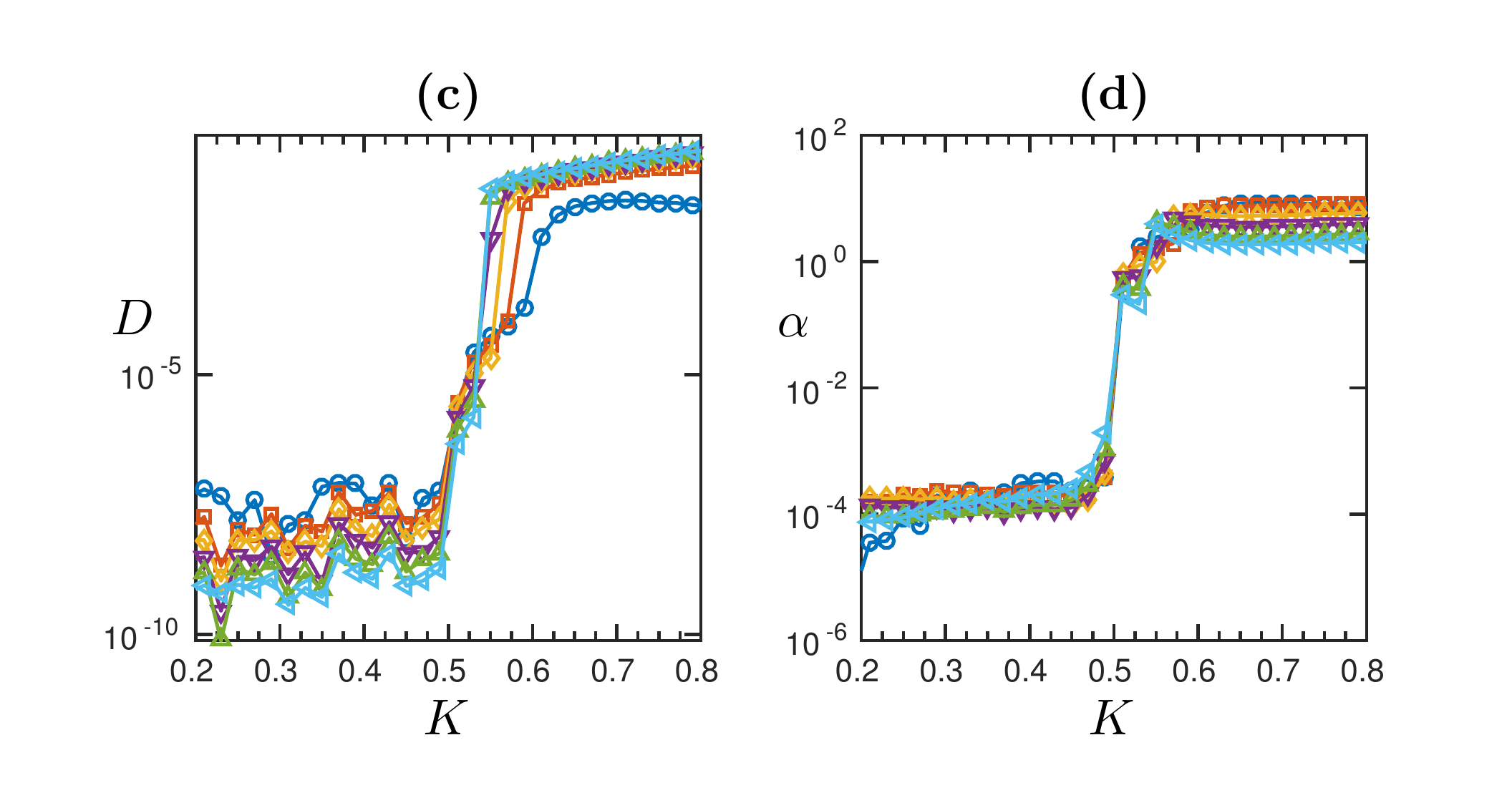}
\caption{(a) Diffusion coefficient $D$ and (b) diffusion exponent $\alpha$ for initial conditions 
close to the stable point ($\phi_0=0,~\sigma=0.1$) with different 
waiting times, $\Delta t=t_f-t_i$. Stability thresholds (resonances) at $K=1$ and $K=0.5$ 
are identified from both the curves. The same curves for $D$ and $\alpha$, respectively, are shown 
in (c) and (d) with logarithmic $y$-axis zooming in around $K=0.5$ to get its non-analytic behavior.}
\label{dc_exp_negK}
\end{figure}

\begin{figure}[t]
\centering
\includegraphics[trim={1.2cm 0.5cm 1.8cm 0.7cm},clip,scale=0.46]{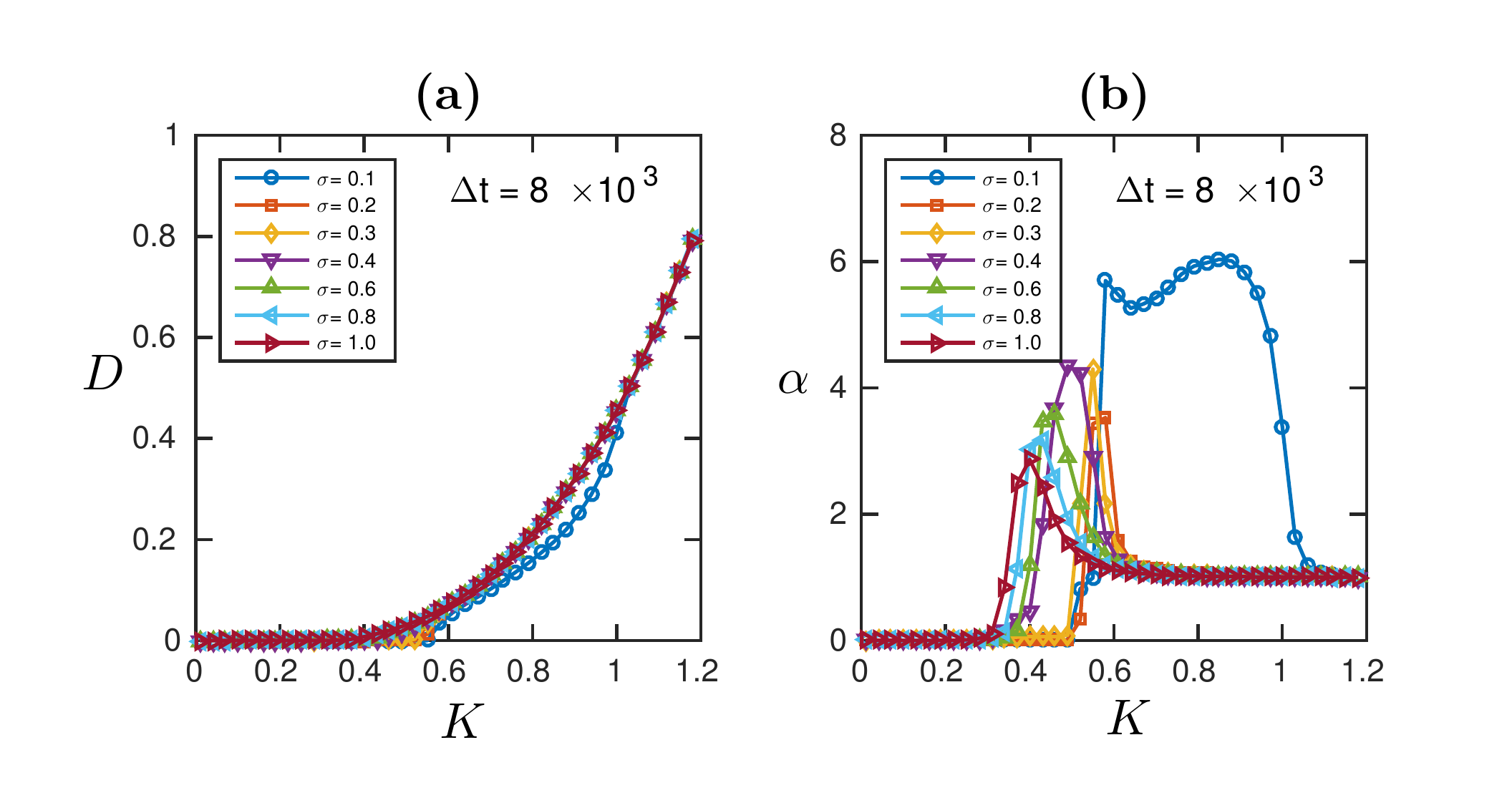}
\includegraphics[trim={1.2cm 0.5cm 1.8cm 0.7cm},clip,scale=0.46]{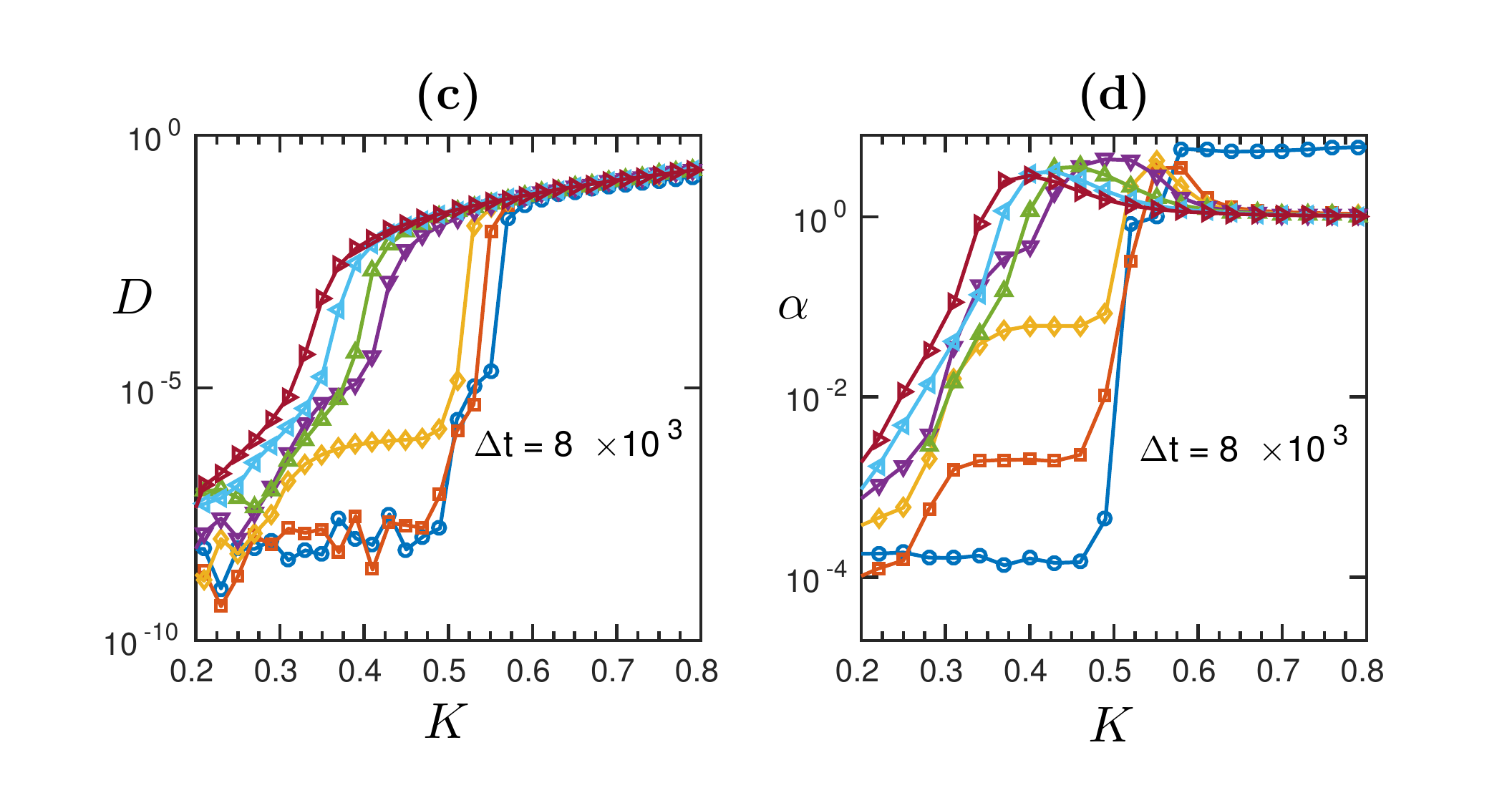}
\caption{(a) Diffusion coefficient $D$ and (b) exponent $\alpha$ at the waiting time $\Delta t=8000$ 
for different values of initial fluctuation (with $\phi_0=0$). For small values of $\sigma$, the system 
shows a sharp transition (resonance) at $K=0.5$, while for large $\sigma$, the curves of $\alpha$ and 
$D$ become smooth functions of $K$. The behavior of (c) $D$ and (d) $\alpha$ around
$K=0.5$ with logarithmic scale in $y$-axis.
}
\label{d_alpha_sd}
\end{figure}

\section{Linear stability and resonances}
\label{ls}
To address the linear stability regime numerically, we consider a system in which all
the initial values of $\phi$'s are taken from 
a Gaussian ensemble with $\phi_0=0$.
First, we consider 
small initial fluctuations: the standard deviation of the Gaussian ensemble is chosen 
to be $\sigma=0.1$. The numerically evaluated diffusion coefficient $D$ and exponent $\alpha$ are shown in Fig.~\ref{dc_exp_negK}. 
In both plots, we find two kinks, respectively at $K=0.5$ and $K=1$. The positions of these two points
do not significantly scale as a function of time.


We can characterize the different 
regions in terms of the behavior of $D$ and $\alpha$ as a function of the waiting time $\Delta t$.
Recall that $D$ and $\alpha$ are calculated by fitting $\langle p^2(t)\rangle$ with linear and power-law 
function of time, respectively. We thus have the relations $\langle p^2(t)\rangle=Dt$ and $\langle p^2(t)\rangle=At^{\alpha}$, that
lead to $D(t)=At^{\alpha-1}$. Therefore it is clear that $D$ does not depend on time for $\alpha=1$.
As shown in Fig.~\ref{dc_exp_negK}(a), for $K>1$ 
the diffusion coefficient $D$ has a non-zero value and all the curves for different time ranges merge 
to each other. Accordingly, as shown in Fig.~\ref{dc_exp_negK}(b), the exponent $\alpha$ converges to $1$ for all the curves. 
Both results indicate that
in this region the system is diffusive. On the other hand, for $K<0.5$, both $D$ and $\alpha$ acquire zero value, 
and the system is localized.
To highlight the behavior of $D$ and $\alpha$ at $K=0.5$, we show the 
same curves around $K=0.5$ with logarithmic scale along vertical axis. From Figs.~\ref{dc_exp_negK}(c,d), 
we observe that both $D$ and $\alpha$ jump from nearly zero values to non-zero one for all the curves 
at $K=0.5$.
For $0.5<K<1$, the exponent $\alpha$ 
decreases with waiting time with two kinks at the
two ends. 
The kink at the point $K=1$ becomes smoother as a function of the waiting time. 
In contrast, the transition to the localized regime at $K=0.5$ is signaled by a kink in $\alpha$ that 
becomes sharper as a function of time.

The numerically determined stability diagram can be understood within the linear analysis presented in Sec.~\ref{qe}. 
As shown there, the points $K=1$ and $K=0.5$ correspond respectively to the 
stability thresholds of the $2^{\rm nd}$ and $4^{\rm th}$ order couplings. For $K>1$, the system is linearly unstable and shows 
a diffusive behavior at all times. In contrast, for $K<0.5$, the system is stable to both quadratic and quartic 
perturbations and is thus localized for all the numerically achievable times. In the intermediate regime ($0.5<K<1$) 
it is super-diffusive for small waiting times, but becomes diffusive for longer waiting times.

As explained in Sec.~\ref{qe}, higher order terms can give rise to resonances at lower values of $K$. 
(For example the $k=6$ coupling becomes unstable at $K=0.25$). To observe these resonances, we repeat
the same calculations with larger values of the initial fluctuations $\sigma$ (at a fixed waiting time). 
For small $\sigma$, the quartic resonance at $K=0.5$ is clearly visible as a kink in the diffusion exponent $\alpha$ (Fig.~\ref{d_alpha_sd}(b)) 
and also a sudden jump in the diffusion coefficient (Fig.~\ref{d_alpha_sd}(a)). 
In contrast, for large initial fluctuations ($\sigma>0.3$), the higher order resonances come into play, and $\alpha$ 
becomes a smooth functions of $K$. We can find a more clear picture how the jump 
in $D$ and $\alpha$ at $K=0.5$ become smoother with increasing initial fluctuations in Figs.~\ref{d_alpha_sd}(c) and (d), respectively.
As we now explain, this situation actually reveals a different type of stability.

\section{Marginal localization}
\label{ms}
\begin{figure}[t]
\centering
\includegraphics[trim={1.2cm 0.5cm 1.8cm 0.7cm},clip,scale=0.46]{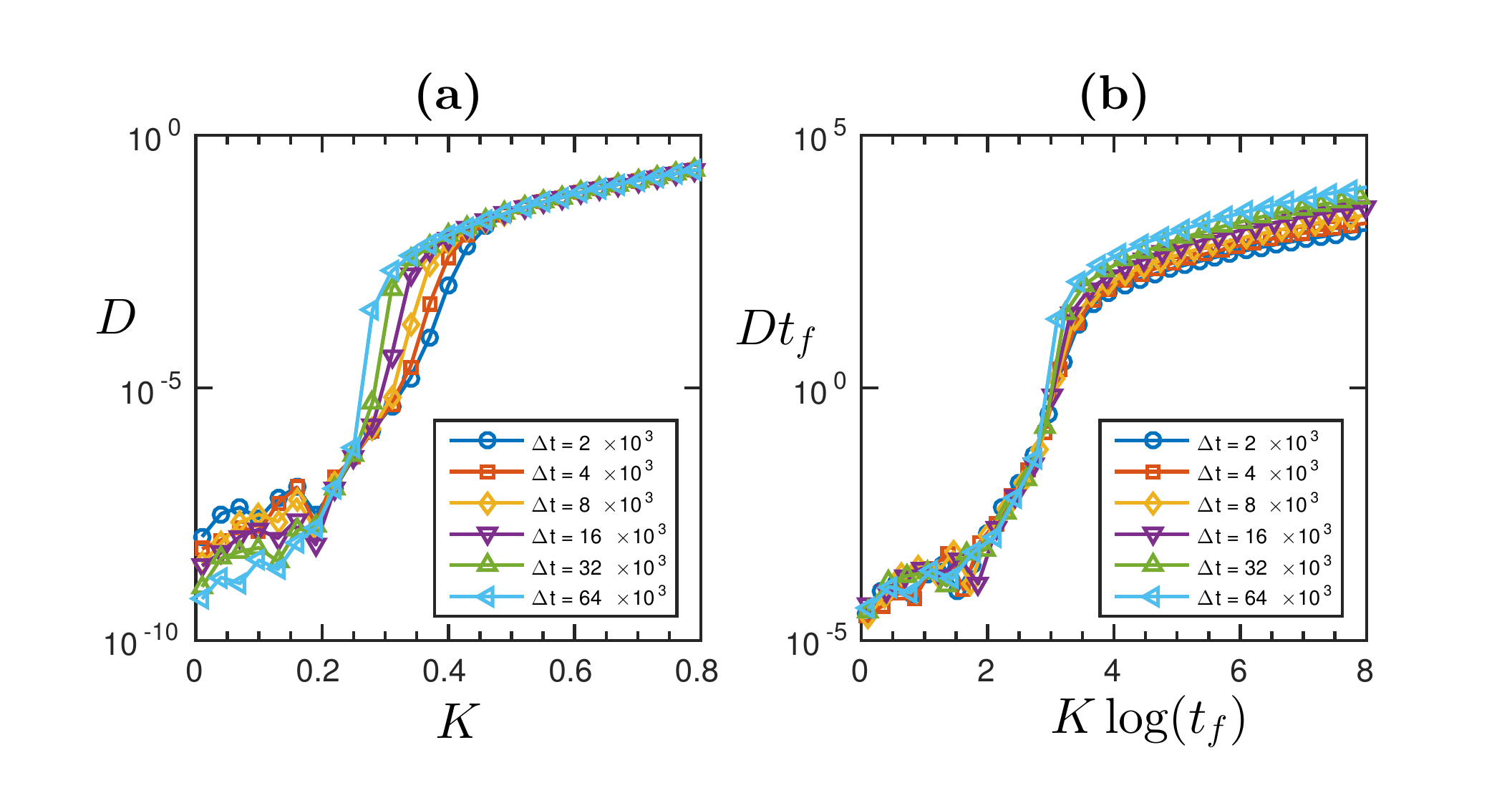}
\caption{(a) Diffusion coefficient $D$ for large initial fluctuations ($\phi_0=0,~\sigma=1.0$),
and different waiting times ($\Delta(t)=t_f-t_i$). (b) For $K\lesssim 3.0/\log t_f$, all the curves can be collapsed by rescaling the $x$ and $y$-axes.}
\label{dc_negK_sd1.0}
\vspace{0.5cm}
\includegraphics[trim={1.2cm 0.4cm 1.8cm 0.7cm},clip,scale=0.46]{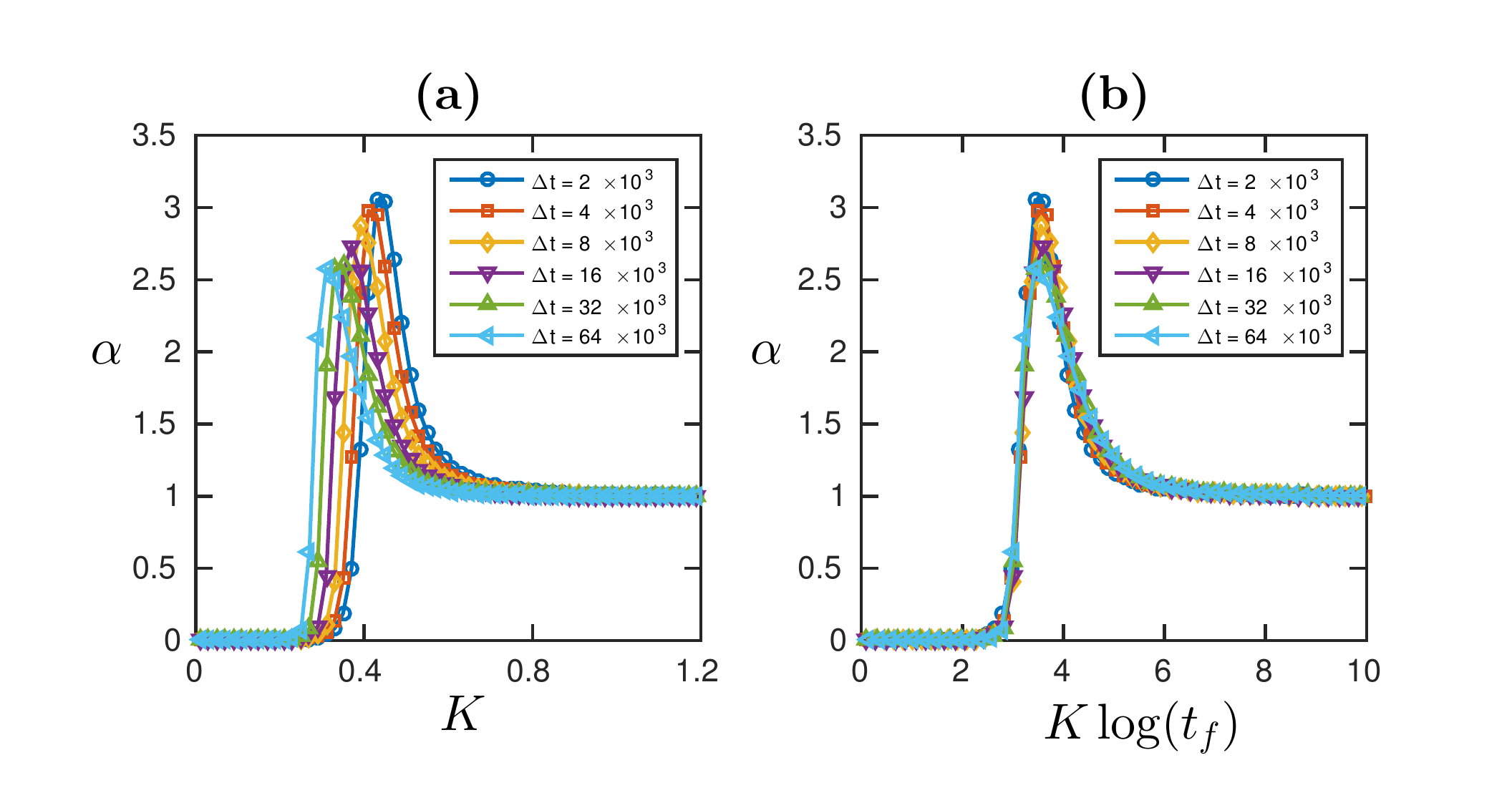}
\caption{(a) Diffusion exponent $\alpha$ for the same initial conditions as in Fig.~\ref{dc_posK}. (b) All the curves collapse when the $x$-axis is rescaled by $\log t_f$. }
\label{exp_negK_sd1.0}
\end{figure}

To investigate the continuous behavior of the diffusion parameters, we now focus on $\sigma=1.0$ for different waiting times. 
In Fig.~\ref{dc_negK_sd1.0}(a), we show that the value of $D$ decreases with time for 
small $K$ and there is a transition point after which $D$ starts increasing with 
time. For $K\gtrsim 1$,  all the curves merge to each other. To better characterize the transition, we apply 
a scaling transformation that leads to a data collapse for small $K$. As shown in Fig.~\ref{dc_negK_sd1.0}(b), this effect can be achieved 
by rescaling  the $x$ and $y$ axes by $\log(t_f)$ and $t_f$ respectively. We observe 
that for $K\lesssim 3.0/\log t_f$ the curves merge to each other, indicating that $D$ is proportional to $1/t_f$.
According to the definition of $D$ in Eq.~(\ref{dc_def}), this finding indicates that the kinetic energy of the system 
does not grow with time for $K\lesssim 3.0/\log t_f$, or equivalently that the system is localized (in momentum space) and does not heat up.

The localized region can be alternatively identified by inspecting the graph of $\alpha$ as a function 
of $K$ for different ranges of time. As shown in Fig.~\ref{exp_negK_sd1.0}(a), these plots show a pronounced peaks, 
whose position moves to lower $K$ as $t_f$ is increased. If we rescale the $x$-axis 
by $\log t_f$, we observe a collapse of all the curves (see Fig.~\ref{exp_negK_sd1.0}(b)).
As expected, the asymptotic behavior of these curves changes at $K_c \approx 3.0 /\log t_f$. 
For $K<K_c$, $\alpha$ decreases as a function of $t_f$, confirming that the system is localized. 
On the other hand, the system is super-diffusive and diffusive for 
$3.0\lesssim K\log t_f\lesssim 6.0$ and $K\gtrsim 6.0/\log t_f$, respectively.  
Interestingly, at $K\approx 3.0/\log(t_f)$ the diffusion exponent $\alpha$ shows a steep increase, 
which sharpens as a function of $t_f$. We conjecture that at very long time,
the model transitions directly from the localized regime ($\alpha=0$) to the super-diffusive one ($\alpha>1$) without showing any sub-diffusive behavior.

As a key result, we find that the the localized region survives until
$t_f = \exp({3.0/K})$ which is exponentially long for small $K = \kappa T$. 
We therefore refer this case as {\it marginal localization}. 
This finding is in line with earlier studies of quantum systems~\cite{abanin15exponentially,goldman15periodically,else17prethermal,abanin17rigorous,chandran16interaction}, 
who showed that the thermalization time is exponentially 
long in the ratio between the driving amplitude and the driving frequency. A similar behavior was observed 
in time-independent classical systems close to an integrable fixed point as well~\cite{pettini90relaxation}.
%

\begin{figure}[t]
\centering
\includegraphics[trim={1.15cm 0.4cm 1.8cm 0.7cm},clip,scale=0.46]{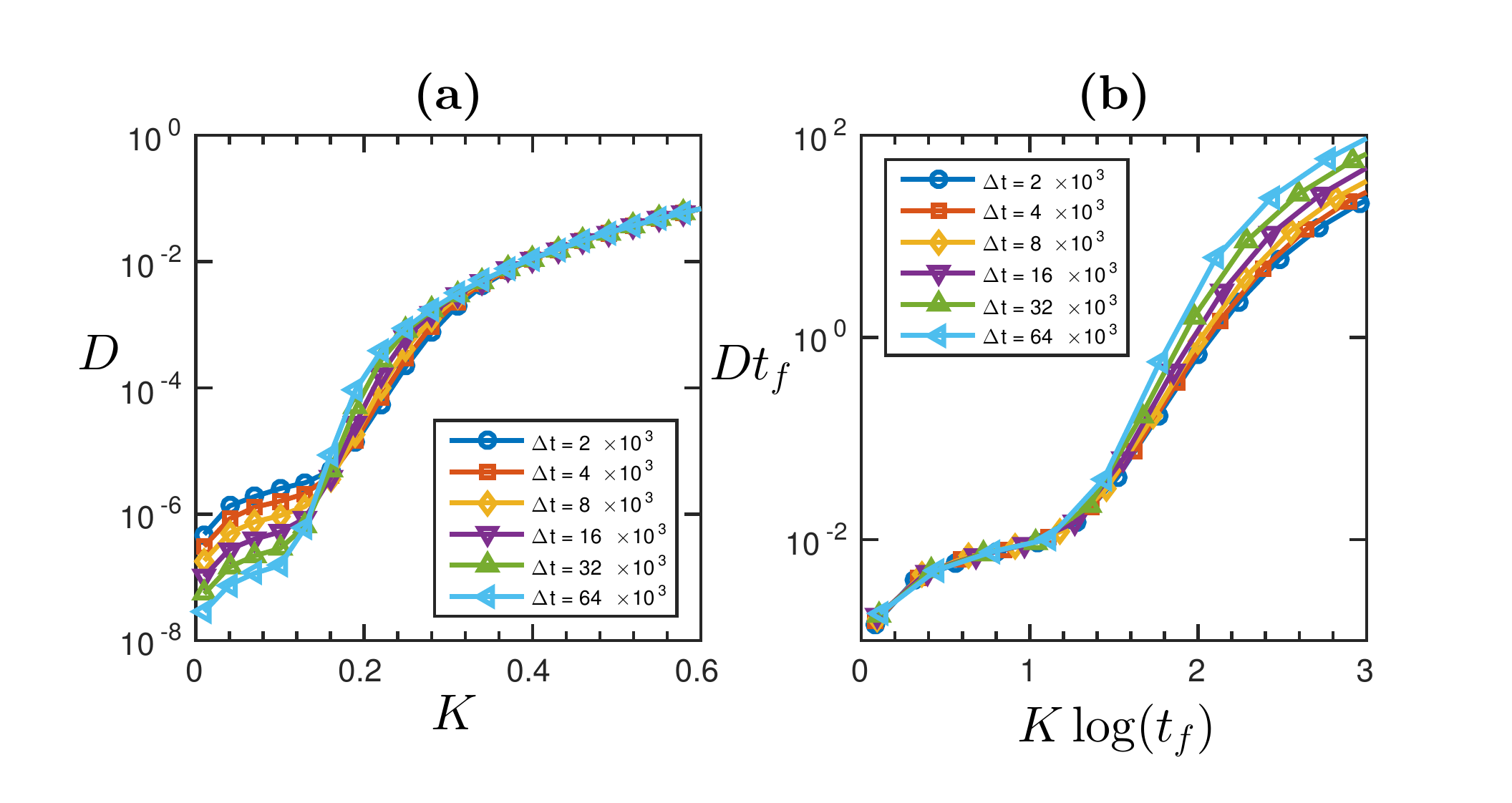}
\caption{(a) Diffusion coefficient $D$ for initial conditions close to the unstable fixed point ($\phi_0=\pi$, $\sigma=0.1$). 
(b) Similarly to Fig.~\ref{dc_negK_sd1.0}(b), all the $D$ curves collapse for $K\lesssim 1.5/\log(t_f)$.}
\label{dc_posK}
\vspace{0.5cm}
\includegraphics[trim={1.1cm 0.5cm 1.8cm 0.6cm},clip,scale=0.46]{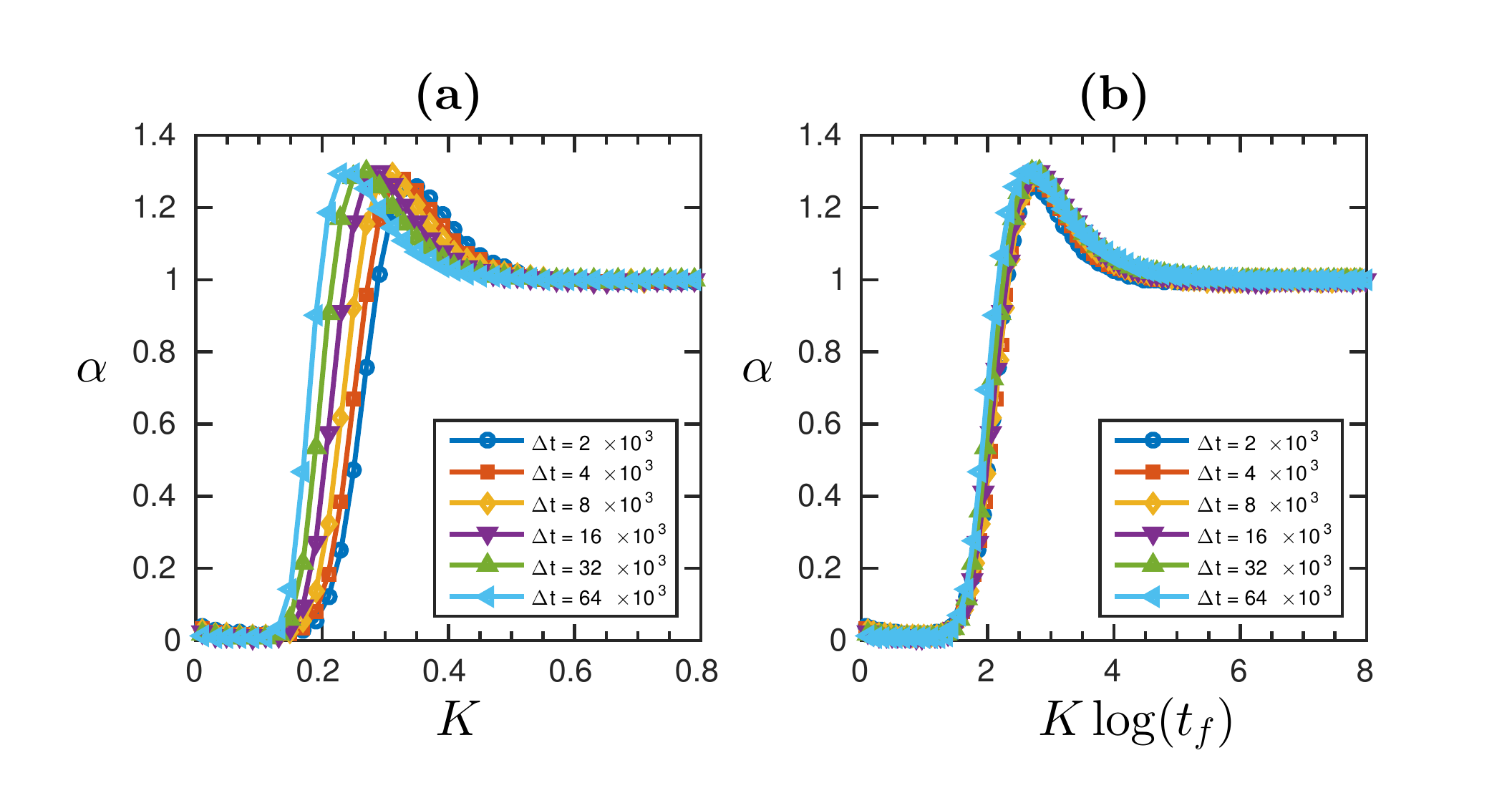}
\caption{(a) Diffusion exponent $\alpha$ for the same initial conditions as in Fig.~\ref{dc_posK}. 
(b) All the curves collapse when the $x$-axis is rescaled by $\log t_f$. 
}
\label{exp_posK}
\end{figure}

The scenario of marginal localization is also achieved by considering initial values of $\phi$'s 
close to the unstable point, $\phi_0=\pi$, even when the initial fluctuations are small ($\sigma=0.1$).
The corresponding results for $D$ and $\alpha$ are shown in Figs.~\ref{dc_posK} and \ref{exp_posK}, respectively. 
The behavior of both the observables is similar to the previous case of marginal localization, although, 
the ranges of the different regimes are bit different compared to the previous case. 
In this case the system is localized for $K\lesssim 1.5/\log t_f$, whereas it becomes super-diffusive 
and diffusive for $1.5\lesssim K\log t_f\lesssim 6$ and $K\gtrsim 6/\log t_f$ respectively.
The transition points differ from the previous case, indicating that the exact 
location of the phase boundaries of the model depend on the specific choice of the initial conditions, 
although the scaling behavior is unchanged.


\section{Wave decomposition}
\label{msr}
The localized and delocalized regimes differ not only in the time evolution of $\langle p^2\rangle$, but 
also in the energy distribution among the Fourier components. To show this effect, we compute 
$\langle|P_q|^2\rangle$ as a function of $q$ inside the different phases. 
Here $P_q$ is the wave component of the momentum of the rotors and is formally defined in Sec.~\ref{qe}.

\begin{figure*}[t]
\centering
\includegraphics[height=4.5cm]{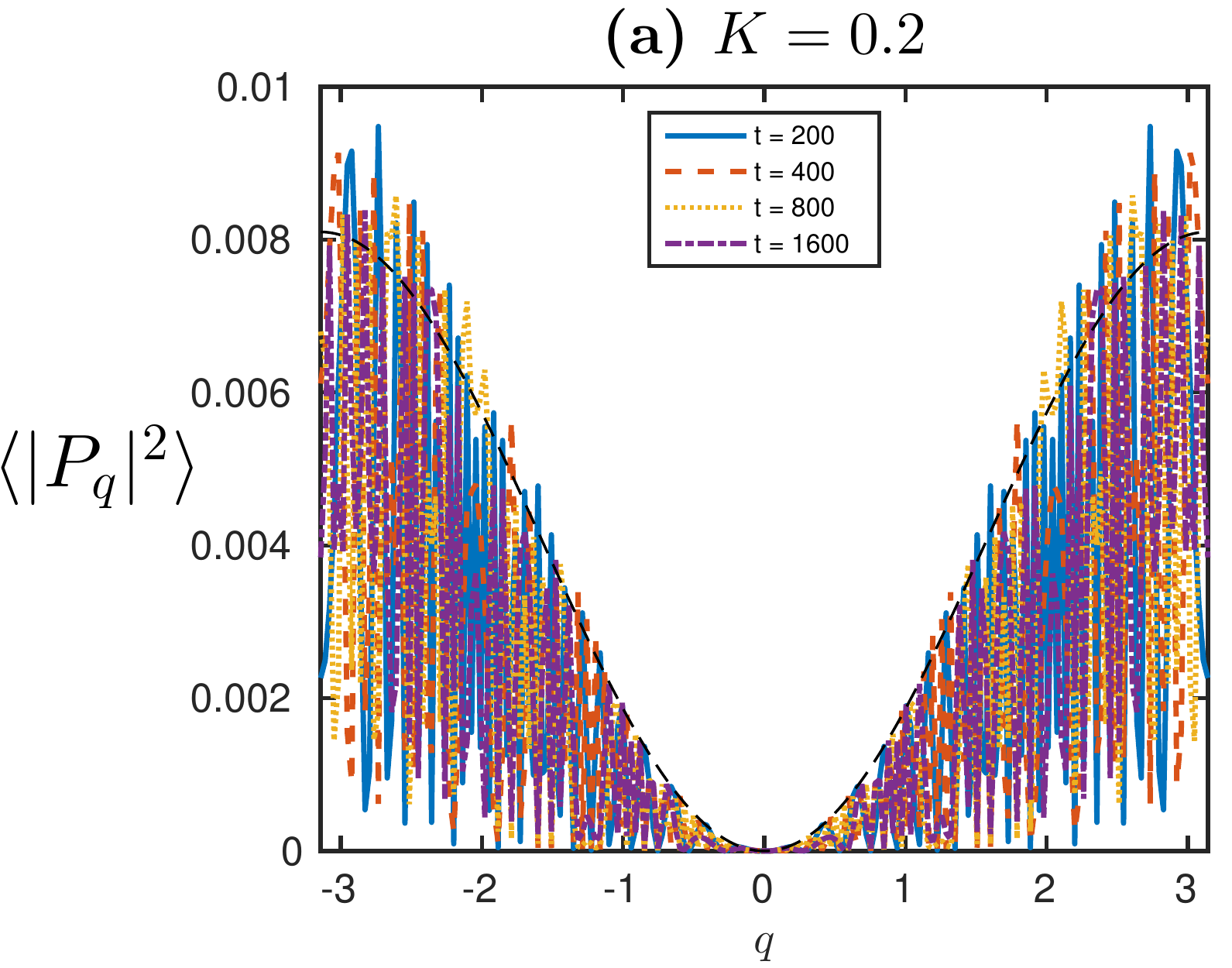}
\includegraphics[height=4.5cm]{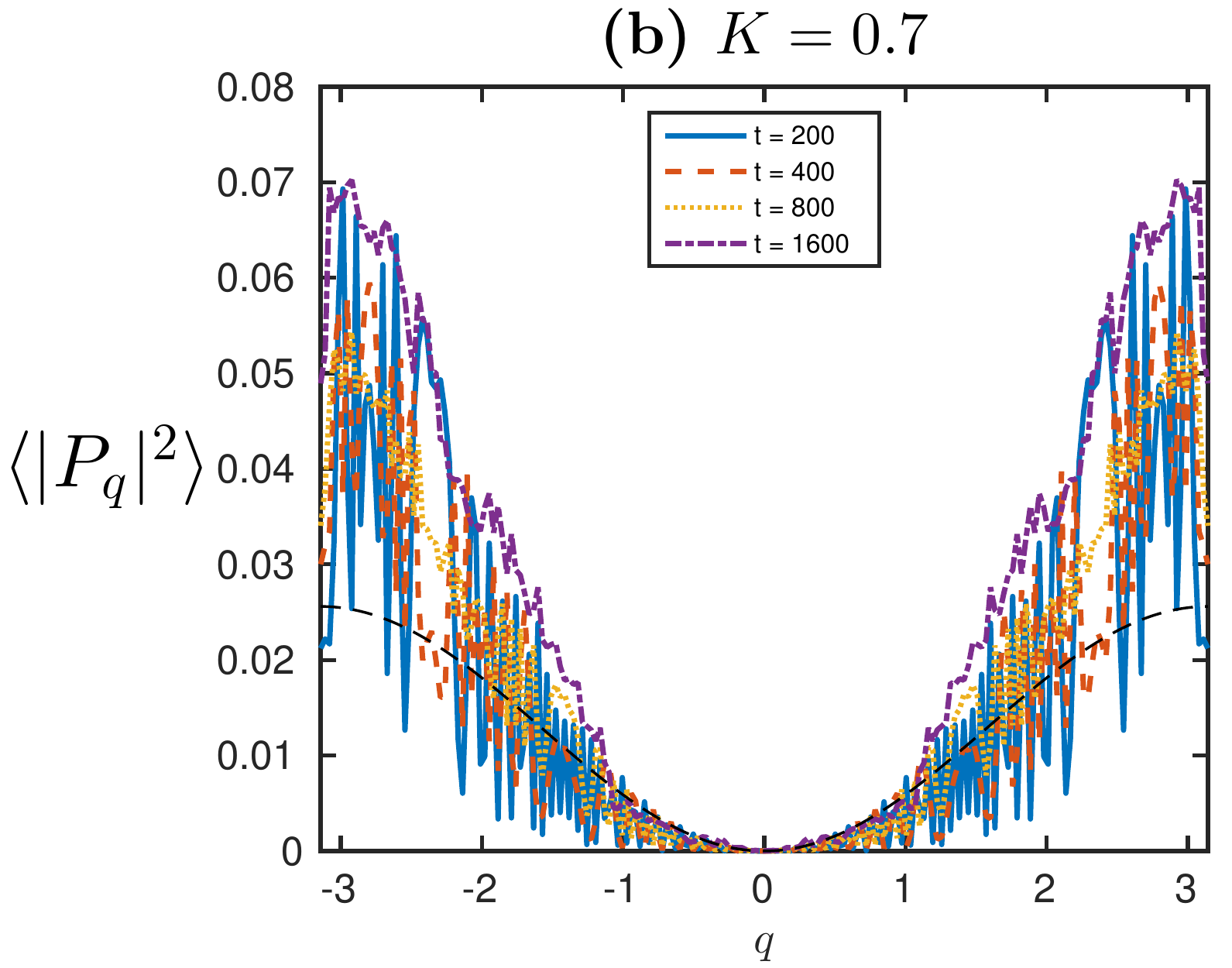}
\includegraphics[height=4.5cm]{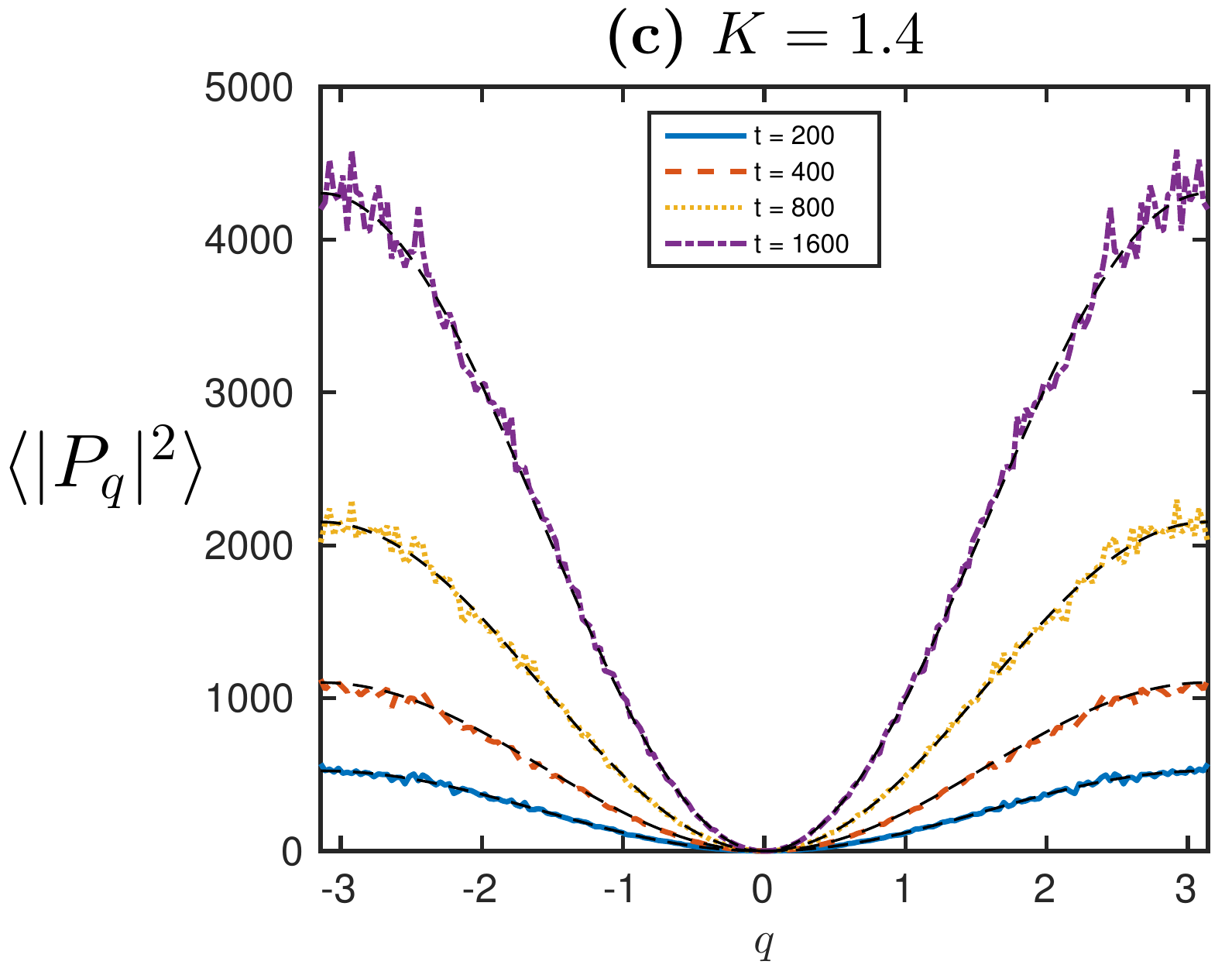}

\caption{Linear stability ($\phi_0=0$, $\sigma=0.1$): Distribution of the kinetic energy over the $q$ modes, for different values 
of the stroboscopic time $t$. The (a-c) subplots represent the stable ($K=0.2$), (b) intermediate ($K=0.7$), and diffusive ($K=1.4$) 
regimes, respectively. The dashed line in each plot 
corresponds to the function $A(q)\propto t\sin^2(q/2)$, which fits well 
the numerical results for the diffusive case ($K=1.4$).}
\label{mom_negK}
\end{figure*}

As before, we divide our discussion between initial conditions close to the stable point ($\phi_0=0$), 
and to the unstable one ($\phi_0=\pi$). The results of the former case are shown in Fig.~\ref{mom_negK}, where we
numerically calculate $\langle|P_q|^2\rangle$  as a function of $q$ for 
$K=0.2, 0.7$ and $1.4$, which correspond to localized, super-diffusive and diffusive phases, respectively 
(see Fig.~\ref{dc_exp_negK}). 
In all the plots the momentum modes close to $q=\pm\pi$ are most excited and the
$q=0$ mode remains unexcited. 
The Hamiltonian in Eq.~(\ref{ham_mkr}) is invariant under the translation transformation
$j\rightarrow j+1$ and, therefore, the total momentum of the system is a conserved quantity.
Since all the initial momenta of the rotors are chosen to be zero, the  momentum 
at $q=0$ remains zero even for later time.
In addition, we have also the relation $P_{-q}=P_q^*$ which makes the spectrum symmetric about 
$q=0$.

Let us first consider the diffusive case ($K=1.4$).
In this regime, $\langle p^2\rangle=\sum_q\langle|P_q^2|\rangle$ grows linearly with time. Fig.~ \ref{mom_negK}(c) 
shows that each $q$ mode grows independently
according to $\langle|P_q^2|\rangle\sim t \sin^2(q/2)$, 
where $t$ is the stroboscopic time. Therefore, from the definition of $|P_q|^2$, we obtain a $q$-dependent 
diffusion coefficient $D(q)\propto \sin^2(q/2)$. We repeated the same analysis 
with initial angles of rotors chosen from another random distribution, but the behavior of the energy per mode
remains intact indicating the form of $D(q)$ is universal in this parameter regime.
Note that in a thermal state, $\langle P_q^2\rangle$ does not depend on $q$. Thus, our findings indicate that the system does not thermalize, 
even at asymptotically long times.
This happens due to the absence of upper bound of energy in the spectrum.
The kinetic energy of the system increases unboundedly with time, which indicates that
it acquires infinite value at infinite temperature ensemble. As a result, the thermalization can not 
occur at any finite time.

\begin{figure*}[t]
\centering
\includegraphics[height=4.5cm]{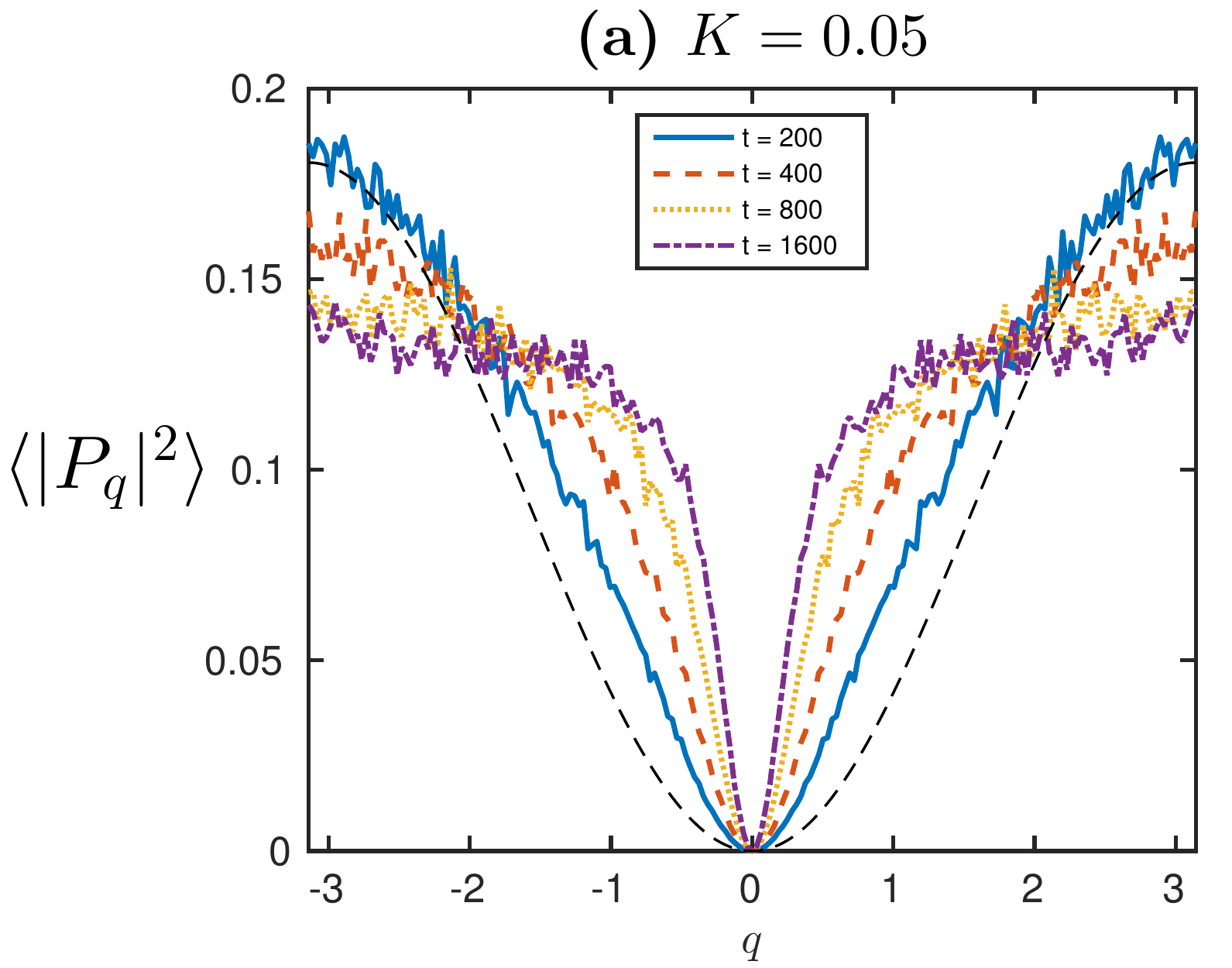}
\includegraphics[height=4.5cm]{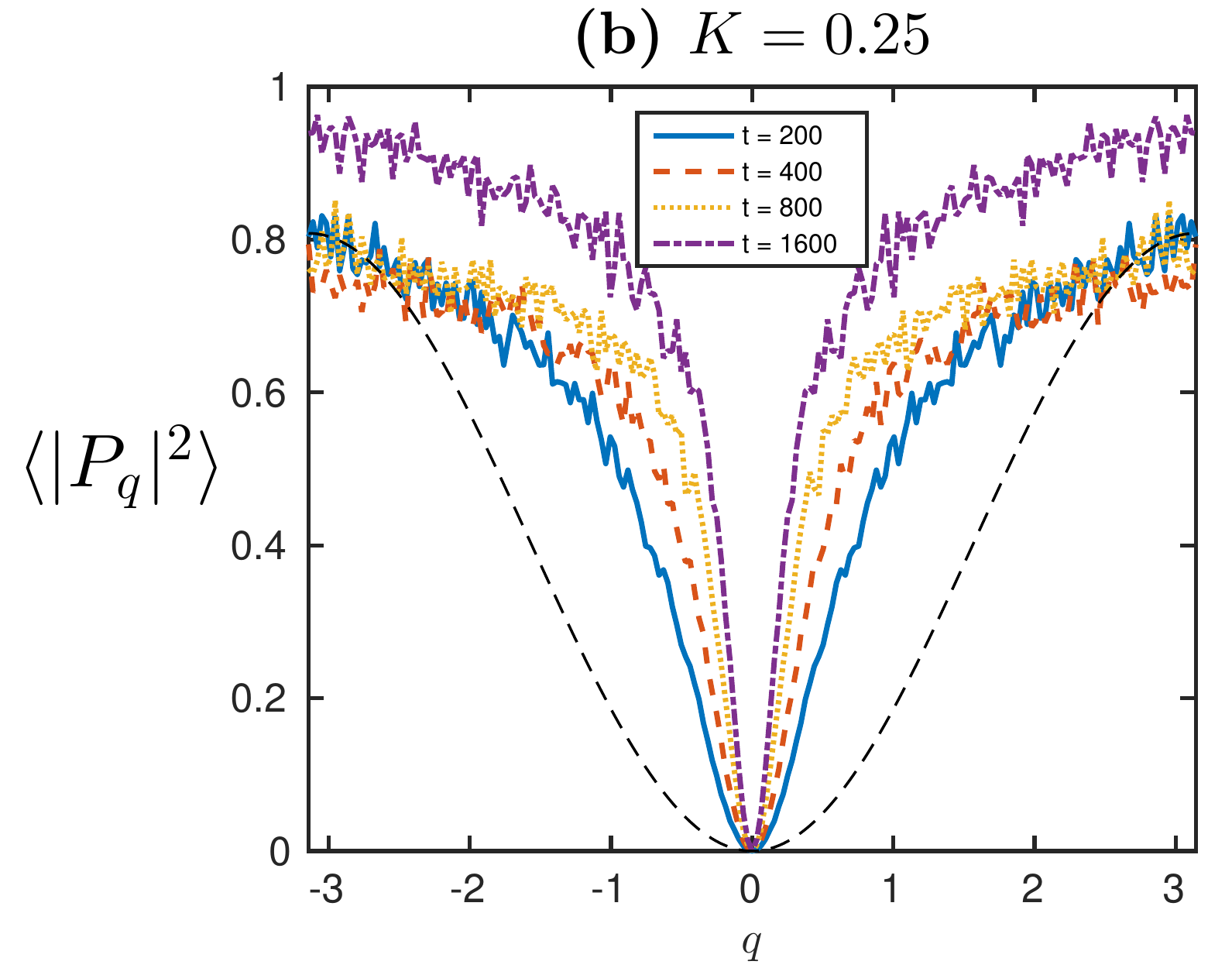}
\includegraphics[height=4.5cm]{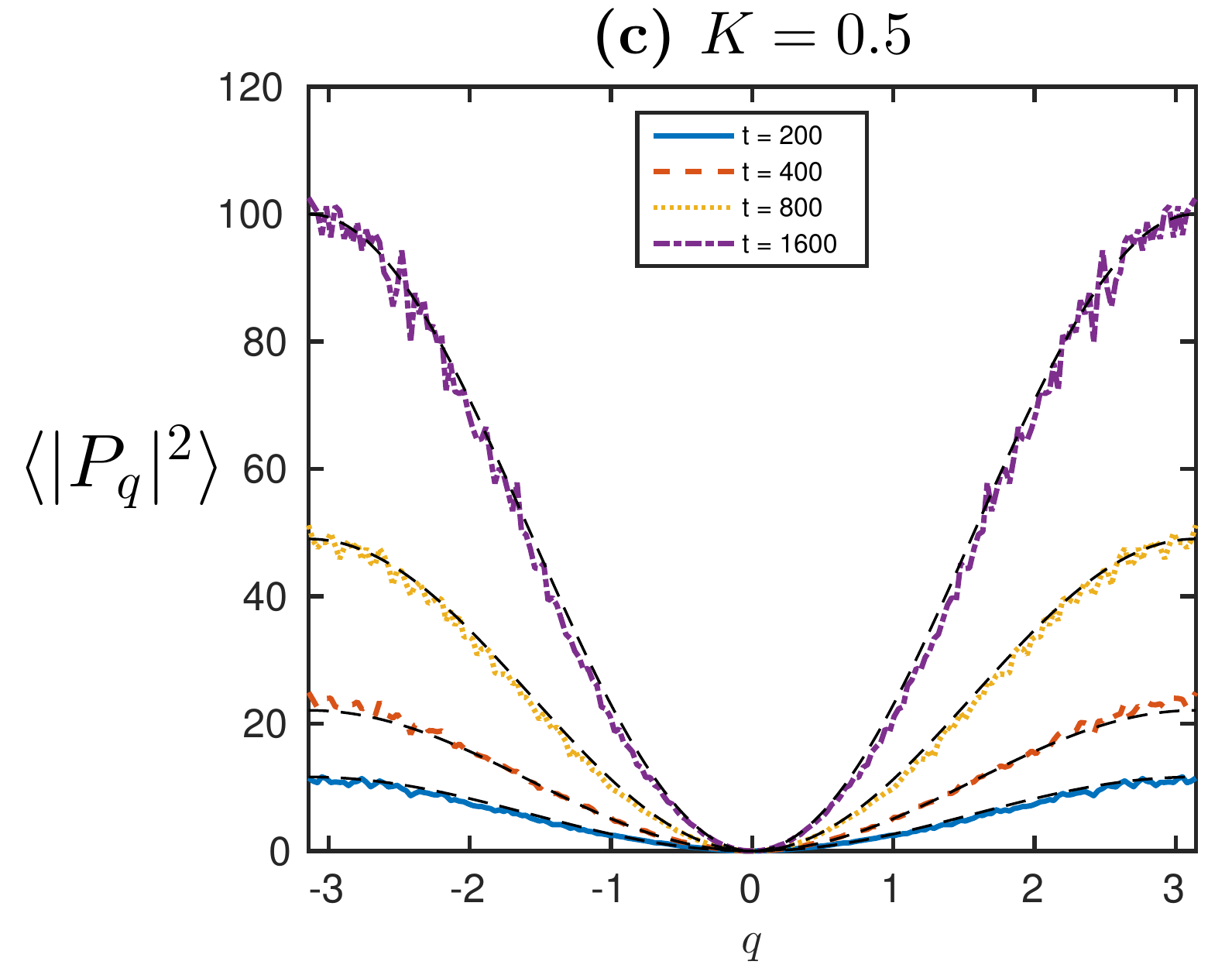}

\caption{Marginal localization ($\phi_0=\pi$, $\sigma=0.1$) : Distribution of the kinetic energy 
over the $q$ modes, for different values of the stroboscopic time $t$. The (a-c) subplots represent 
the marginally localized ($K=0.05$), (b) super-diffusive ($K=0.25$), and diffusive ($K=0.5$) regimes, 
respectively. The dashed line is a plot of the function 
$A(q)\propto t\sin^2(q/2)$ which fits well the numerical results 
for the diffusive case ($K=0.5$).}
\label{mom_posK}
\end{figure*}

We next consider the localized regime ($K=0.2$).
As shown in Fig.~\ref{mom_negK}(a),
at a given time $\langle|P_q|^2\rangle$ randomly oscillates as a function of $q$. The range of the oscillations does not depend on time. 
As discussed in Sec.~\ref{qe}, the Hamiltonian 
in Eq.~(\ref{ham_mkr}) can be expanded around the stable point $\phi_0=0$ and it provides a collection of decoupled 
harmonic oscillators in the Fourier space. Therefore, for this case, the energy oscillates 
between the position and momentum of the $q$ mode, $\phi_q$ and $P_q$, for each $q$, but the total energy remains constant with time. 
This phenomenon is reflected in the oscillations of $\langle|P_q^2|\rangle$ as a function of $q$ at a fixed time. As expected from the linear model,
we do not observe energy exchange among the wave modes. 

In the intermediate regime, at $K=0.7$, the average energy per mode shows oscillations for shorter times, 
while the oscillations die out as time increases (see Fig.~\ref{mom_negK}(b)). At long times, the modes around $q=\pi$ feel 
the $4^{\rm th}$ order resonance described in Sec.~\ref{qe}, and the shape of $\langle|P_q^2|\rangle$ starts deviating from the simple $\sin^2(q/2)$ shape. 
This effect is associated with the super-diffusive behavior described in Sec.~\ref{ls}.

We now discuss a system that shows marginal localization, by considering initial conditions close to the unstable 
point ($\phi_0=\pi$). Following the same strategy as before, we numerically 
calculate the energy distribution for 
the localized, super-diffusive and diffusive regimes (see Fig.~\ref{mom_posK}). 
As shown in Fig.~\ref{mom_posK}(c), the behavior of $\langle|P_q|^2\rangle$ in the diffusive regime 
is very similar to the previous case.


In the marginally localized regime ($K=0.05$), the wave modes 
close to $\pm\pi$ get de-excited, whereas the modes near $q=0$ become excited as the number of kicks is increased. 
This indicates that the excited modes transfer energy to relatively less excited modes, but 
the system does not absorb energy from the driving. As a result, $\langle p^2\rangle$ 
does not grow with time in this regime leading to vanishing values of $D$ and $\alpha$ (see Figs.~\ref{dc_posK} and \ref{exp_posK}). 
By extrapolating Fig.~\ref{mom_posK}(a) to asymptotically long times, one finds that $\langle|P_q^2|\rangle$ 
tends to a constant function (with a singularity at $q=0$).
This finding implies that, although, the system eventually 
becomes diffusive at exponentially long time (see discussion in Sec.~\ref{ms}), 
it prethermalizes to a finite effective temperature at
intermediate time
\footnote{The phenomenon of prethermalization is also observed in spinless fermion models with a weak 
integrability-breaking term after a quantum quench~\cite{bertini15prethermalization}. In this case, the prethermalization time shows a power law 
variation with interaction term, which is in contrast to our observation of exponentially long localization 
time for a periodically driven model.}
This finding is in agreement with recent studies of quantum cases 
~\cite{bukov15prethermal,kuwahara16floquet,canovi16stroboscopic,weidinger17floquet,zeng17prethermal,else17prethermal,abanin17effective},
where the authors showed that the driven quantum systems can prethermalize to a thermal steady state at some 
intermediate time scale. 

The intermediate regime ($K=0.25$) 
is shown in Fig.~\ref{mom_posK}(b). Its behavior 
is also intermediate between the localized and diffusive phases. 
For short times, the system is in the localized regime and transfers energy from large $q$ to small $q$. 
In contrast, for longer times the modes close to $q=\pi$ become excited, leading to a super-diffusive dynamics.



\section{Summary and conclusions}
\label{conclusion}
In this paper, we revisited the problem of stability and thermalization of 
periodically driven classical many-body system. In quantum systems, the eigenstate thermalization hypothesis (ETH) 
states that the system should thermalize at long times, but it does not make any prediction about the dynamics of the system at 
intermediate time scales. For instance, earlier studies showed that the system can reach an intermediate long-lived 
pre-thermalized state before approaching to the infinite temperature 
ensemble. We addressed the classical analogue of this question by investigating the finite-time dynamics of 
a classical many-body kicked rotor, whose dynamics is given by a
generalization of the ``standard map" (see Eq.~(\ref{cd_equs1})).


When the initial angles of the rotors are chosen in vicinity of the stable point, 
we observed two sharp transitions, whose position does not scale with the waiting time.
The first transition ($K=1$) is observed only at short times, and corresponds to an exceptional 
point of the linear dynamics, where the fixed point becomes unstable. The second transition ($K=0.5$) 
becomes sharper as a function of time, and is due to quartic terms, which can lead to energy 
absorption in the linear stability regime. The small initial fluctuations suppress the effects of higher-order terms, 
and lead to a system that is stable for all numerically achievable times: In this regime the system is nearly integrable 
and never equilibrates~\ct{russomanno12periodic,russomanno2016kibble,
ponte15many,lazarides15fate,ponte15periodically,abanin16theory,agarwal17localization,dumitrescu17logarithmically}.


For large initial fluctuations, the linear analysis breaks down and both the diffusion coefficient $D$ and exponent 
$\alpha$ become continuous functions of $K=\kappa \tau$.
A similar situation appears when the initial angles of the rotors are considered around the unstable point. 
For both the cases, we observe a collapse of the $\alpha$ curves for different 
waiting times by rescaling the $x$-axis with $\log t_f$. This finding indicates 
that the localized phase of the system for small $K$ sustains for an exponentially long time 
and the system is {\it marginally localized}. In other words,
the rate of approach of the system to the diffusive 
phase is logarithmically small predicting the existence of a long-lived prethermal regime.

To analyze the phenomenon of prethermalization, and to study the localized and delocalized 
regimes further, we also considered the Fourier space representation of the momenta
of the rotors. In this connection, we numerically calculated the average fluctuations of 
the wave components in the different regimes.
In the diffusive phase the system absorbs energy from the external driving and 
the average momentum grows linearly in time. Interestingly, the kinetic energy does not 
redistribute evenly among the wave-vectors, but remains proportional to $\sin^2(q/2)$ at all times. 
This observation indicates that although the system is diffusive, it does not thermalize. The reason for this anomaly is found 
in the un-bound nature of the system's spectrum: Because the kinetic energy is proportional to $p^2$, 
its expectation value should be infinite in an infinite temperature ensemble. Clearly, this cannot happen at any finite time.

In the marginally localized regime, the average energy of the wave modes at small $q$ increases with time, while it decreases 
for modes close to $q=\pm\pi$. This indicates 
that energy is transfered from excited to de-excited modes, but the system does not absorb 
energy from external driving. 
The expectation value of $\langle P_q^2\rangle$ tends to a constant (with a sharp dip at $q=0$), suggesting that 
the system is flowing towards a prethermalized state, whose energy is set by the fluctuations of the initial state. 
At exponentially long times, the system escapes from this prethermal state and becomes diffusive. 
The transition between the marginal and diffusive regimes occurs through an intermediate super-diffusive regime, 
where energy absorption and thermalization compete.

\begin{table}
\centering
\begin{tabular}{|m{1.6cm}|m{1.6cm}|m{1.9cm}|m{1.9cm}|m{1cm}|}
\hline
 & Initial conditions & Localized & Diffusive & Figures\\
\hline
Linear stability & $\phi_0=0,~~~~$ $\sigma=0.1$ & $K<0.5$ & $K>0.5$ & 1, 7\\
\hline
\multirow{2}{*}{\parbox{1.8cm}{\begin{flushleft}Marginal localization\end{flushleft}}} & $\phi_0=0,~~~~$  $\sigma=1.0$ & $K\lesssim\frac3{\log(t_f)}$ & $K\gtrsim\frac6{\log(t_f)}$ & 3, 4\\
\cline{2-5}
& $\phi_0=\pi,~~~~$ $\sigma=0.1$ & $K\lesssim\frac{1.5}{\log(t_f)}$ & $K\gtrsim\frac6{\log(t_f)}$ & 5, 6, 8 \\
\hline
\end{tabular}

\caption{Summary of the numerical results of this study (see text for details).}
\end{table}

Our analysis highlights two alternative methods to stabilize periodically driven many-body systems: (i) linear stability, 
induced by the choice of initial conditions close to a fixed point of the classical dynamics, and (ii) marginal localization, 
obtained for weak and/or fast drives (see Table I). In both cases, the system can remain stable for an arbitrarily long time: in the former 
case the instability is suppressed by the closeness to the fixed point, while in the latter it is suppressed (exponentially) 
by the ratio between the driving frequency and the driving amplitude. Our analysis clarifies the nature of many-body 
instabilities observed in the earlier literature. In particular, Refs.~\cite{prosen98time,citro15dynamical,bhadra2018dynamics} considered an initial state with 
(quasi) long range correlations, which can stabilize the system in the linear stability mechanism. 
In contrast, the prethermal states observed by Refs.~\cite{zeng17prethermal,abanin15exponentially,abanin17rigorous, 
else17prethermal,machado17exponentially,dumitrescu17logarithmically,weidinger17floquet}
are protected by a large driving frequency, and thus fall into the category of marginal localization. 
These earlier works considered many-body quantum systems, but the effects that they describe exist in classical systems as well.

\vskip 0.5cm
\noindent {\bf {Acknowledgements}}
\vskip 0.5cm
\noindent We thank Ehud Altman, Nivedita Bhadra, Itzhack Dana, Anatoli Polkovnikov, Francesco Romeo, Angelo Russomanno, and Marco Schiro for many fruitful discussions. 
This work was supported by the Israeli Science Foundation Grant No. 1542/14.

\appendix
\section{Variation of $\langle p^2\rangle$ with time}
\label{appendixt}
\begin{figure}[h]
\centering
\includegraphics[height=4.2cm]{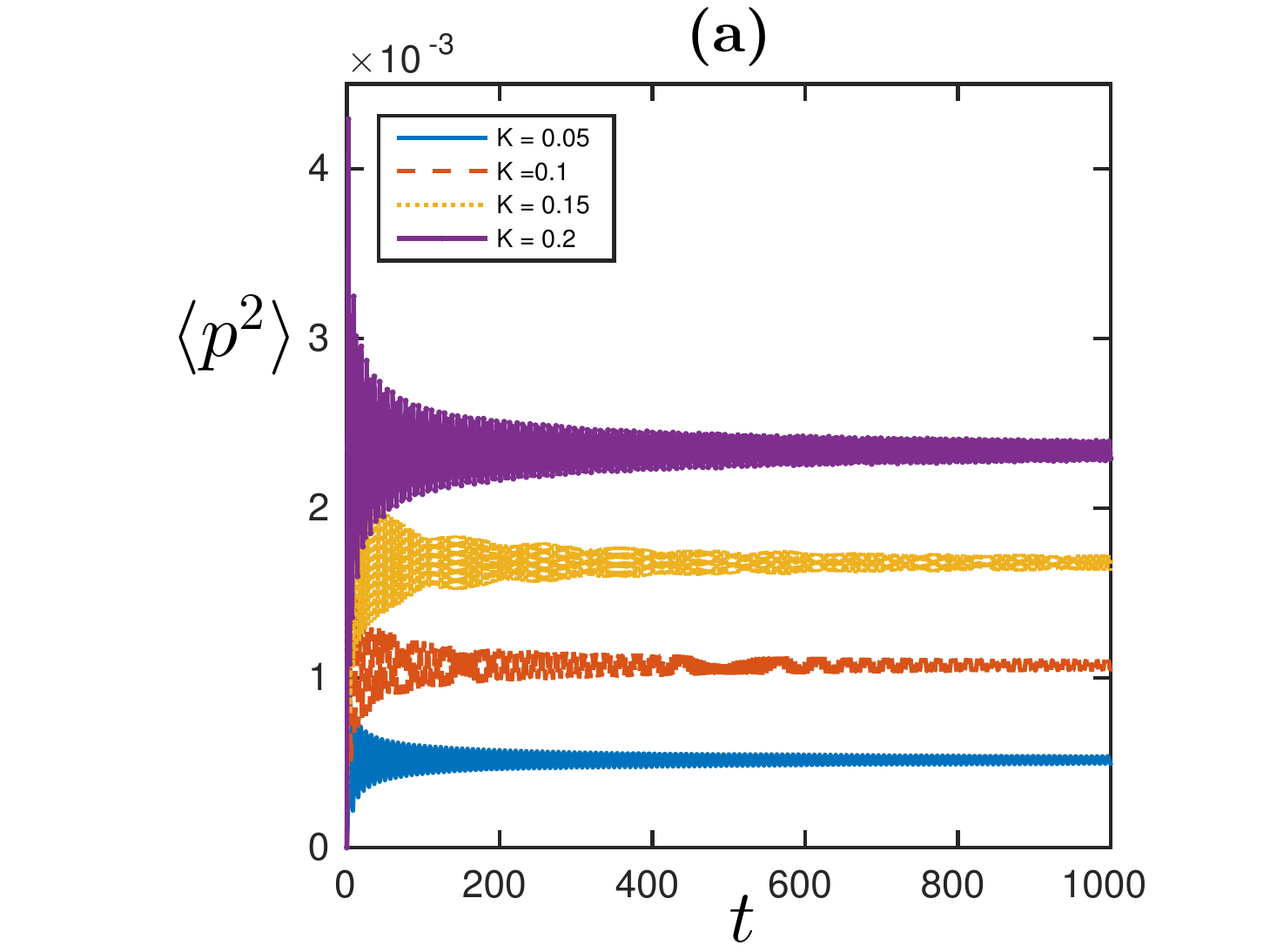}
\hspace{-0.78cm}
\includegraphics[height=4.2cm]{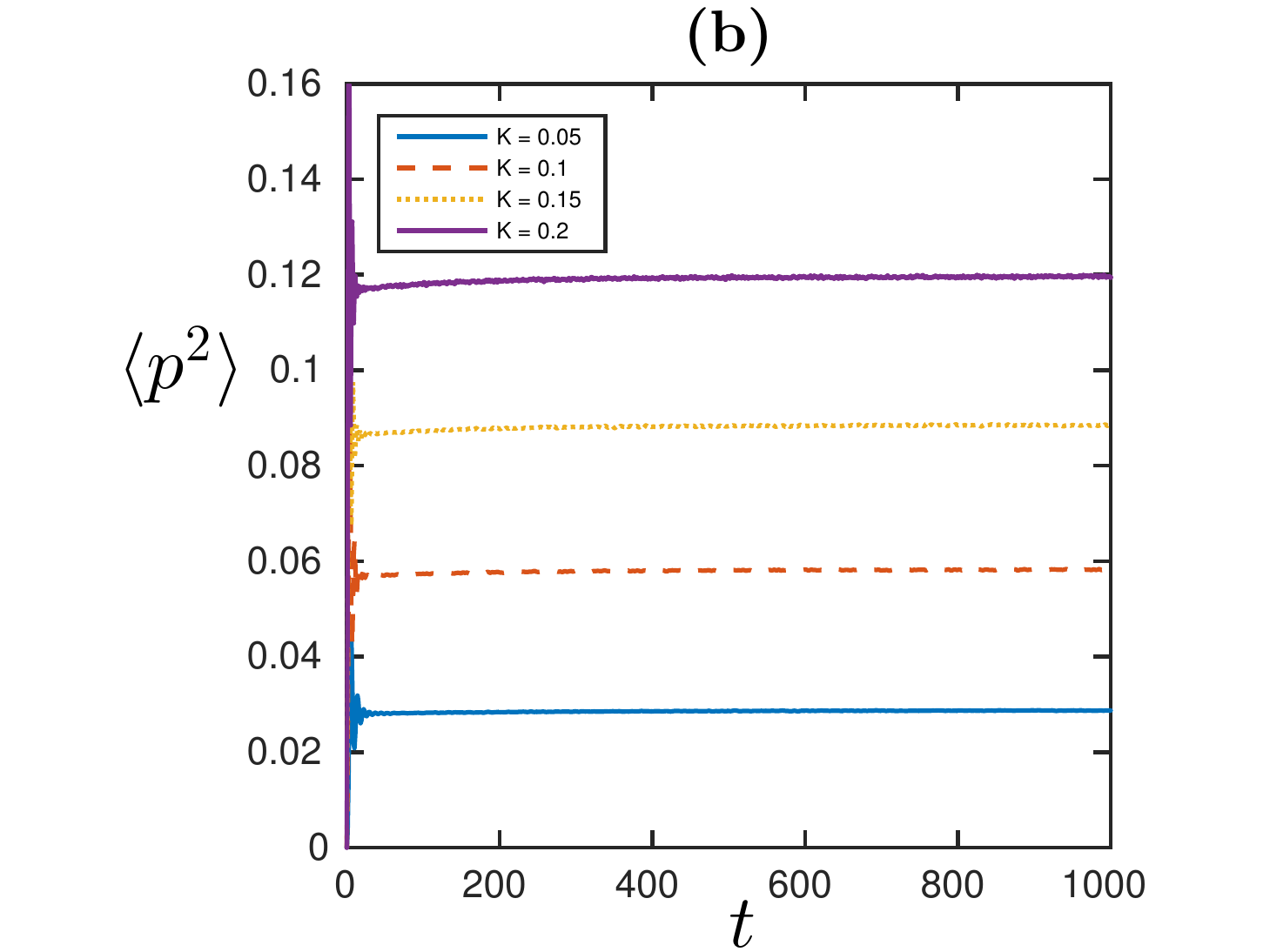}
\includegraphics[height=4.2cm]{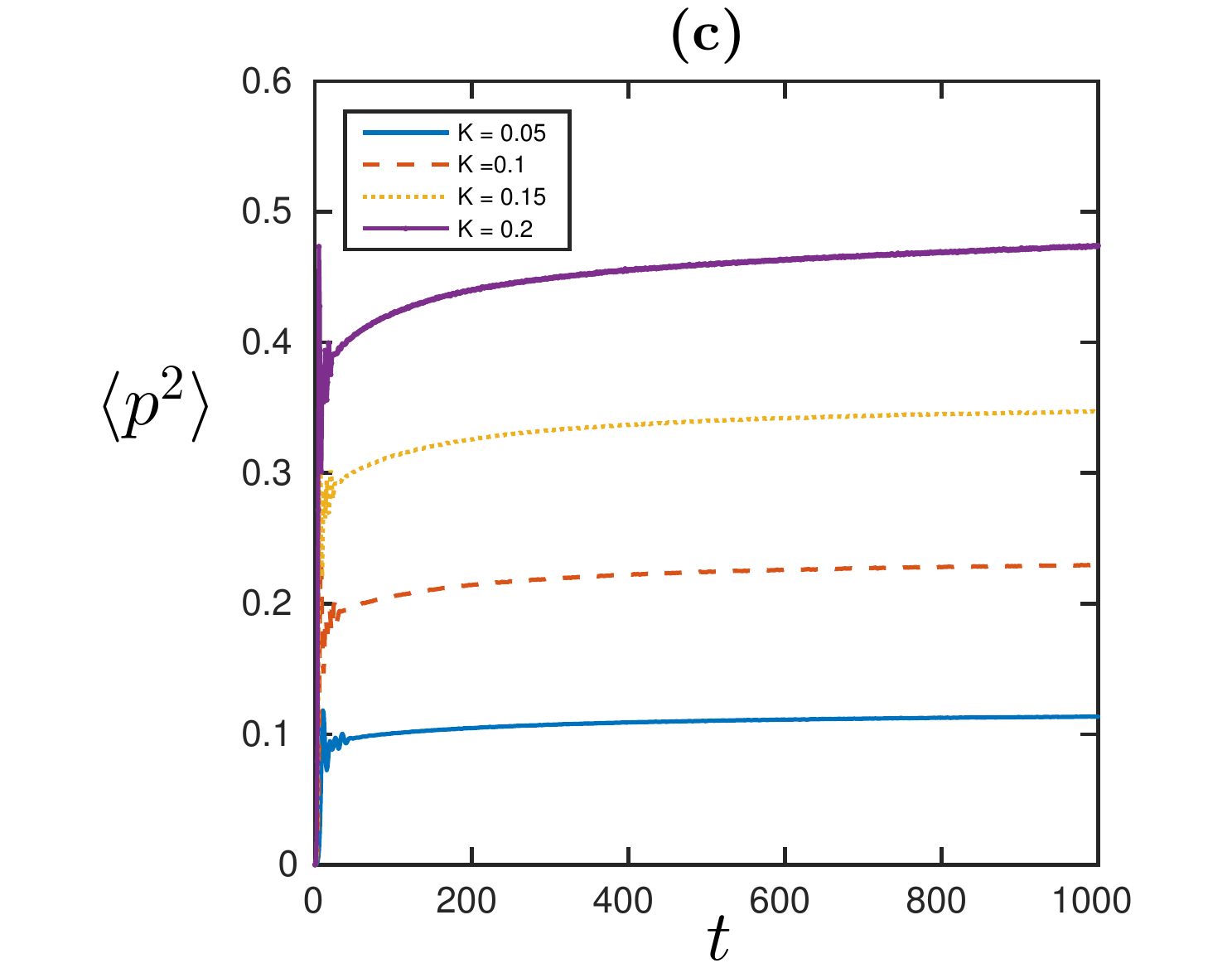}
\caption{$\langle p^2\rangle$ for relatively small $K$ 
as a function of time $t$ for (a)  initial conditions ($\phi_0=0$, $\sigma=0.1$) same as in 
Fig.~\ref{dc_exp_negK}, (b) initial conditions ($\phi_0=0$, $\sigma=1.0$) same as in 
Figs.~\ref{dc_negK_sd1.0} and \ref{exp_negK_sd1.0}, and (b) initial conditions ($\phi_0=\pi$, $\sigma=0.1$) 
same as in Figs.~\ref{dc_posK} and \ref{exp_posK}.}
\label{ee_t_np}
\end{figure}

In this appendix we show typical plots of the time evolution of $\langle p^2\rangle$, which were used to extract the diffusion 
parameters $D$ and $\alpha$. Fig. \ref{ee_t_np}(a) refers to the time evolution in the linearly localized regime. On the other hand, 
Figs.~\ref{ee_t_np} (b) and (c) exhibit the time evolution for the two scenarios of marginally 
localized regime (see Sec.~\ref{ms}).

In all cases, the time evolution initially shows fast oscillations, associated with the equipartition of energy between 
the $\phi$ and $p$ degrees of freedom. For Figs.~\ref{ee_t_np} (b) and (c), these oscillations are followed by a smooth evolution in time, 
which is the focus of the present article. In contrast, the oscillations in Fig. \ref{ee_t_np}(a) sustain for very long time, since 
the initial conditions are chosen close to the stable point ($\phi_0=0$). However, the amplitude of oscillations decreases in time.
For this reason, we extract $D$ and $\alpha$ from the time evolution of $\langle p^2\rangle$ at times $t>t_i=1000$ 
only. We checked that our results are independent on the specific choice of $t_i$.

\section{Effect of finite system size and error due to random initial conditions}
\label{apendixa}
\begin{figure}[h]
\centering
\includegraphics[trim={1.2cm 0.5cm 1.8cm 0.7cm},clip,scale=0.46]{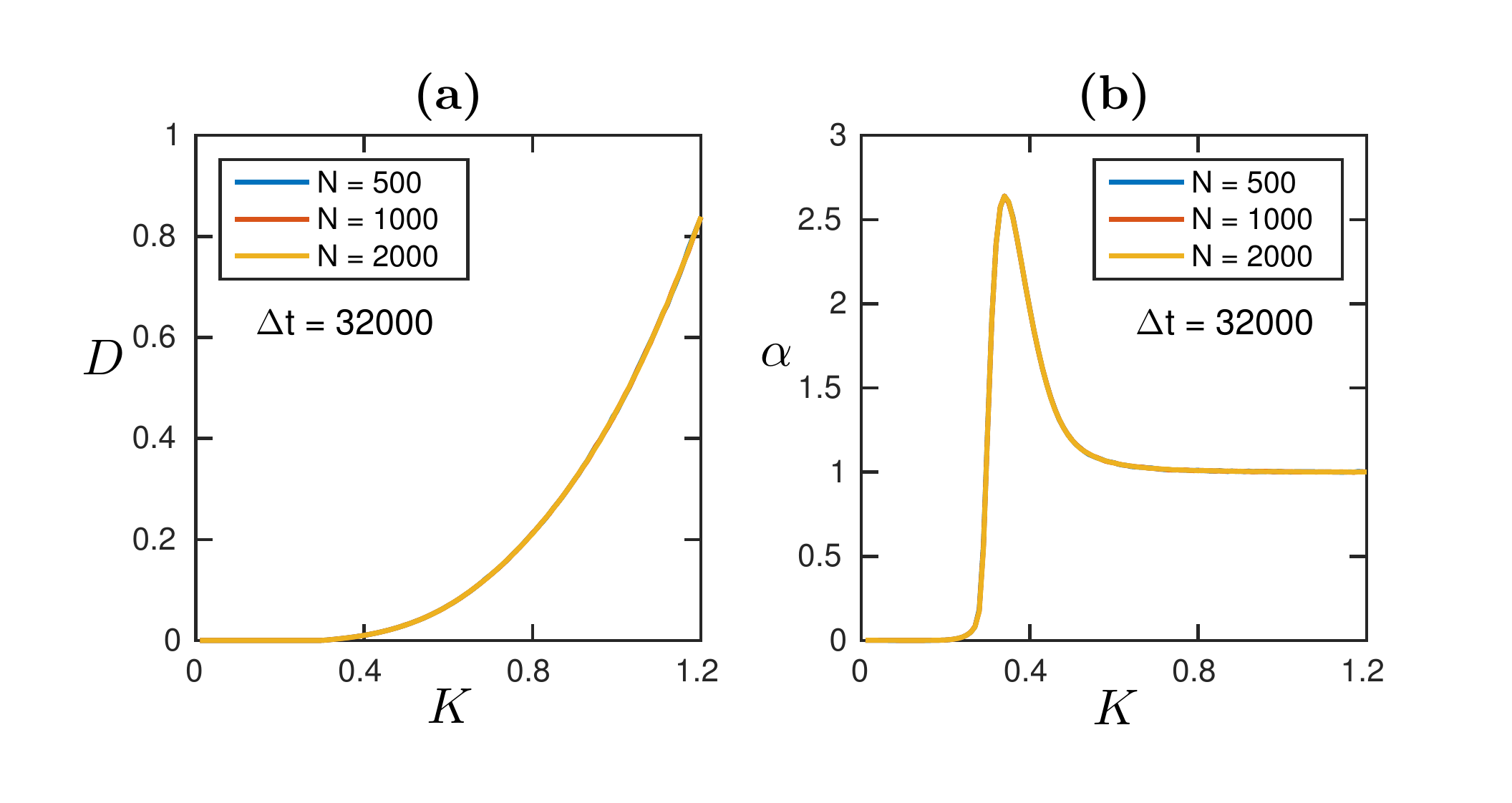}
\caption{(a) Diffusion coefficient $D$ and (b) diffusion exponent $\alpha$ for different system sizes ($N$) at
waiting time $\Delta t=32000$. The plot shows that finite size effects are negligible. Here $\phi_0=0$ and $\sigma=1.0$.}
\label{d_alpha_n}
\end{figure}

In this Appendix we investigate whether our numerical results have finite size effects. 
In addition, we also analyze the error in our data considering the plots 
as shown in Figs.~\ref{dc_posK} and \ref{exp_posK}.
For the first check we calculate both $D$ and $\alpha$ for three different system 
sizes with one waiting time. As shown in Fig.~\ref{d_alpha_n}, all the curves for different 
system sizes perfectly overlap each other exhibiting that our results are free from finite system size 
effects. However there exists the effect of finite time for which we have provided a clear picture 
in the main text considering different times and initial conditions.

\begin{figure}
\centering
\includegraphics[trim={1.2cm 0.5cm 1.8cm 0.7cm},clip,scale=0.46]{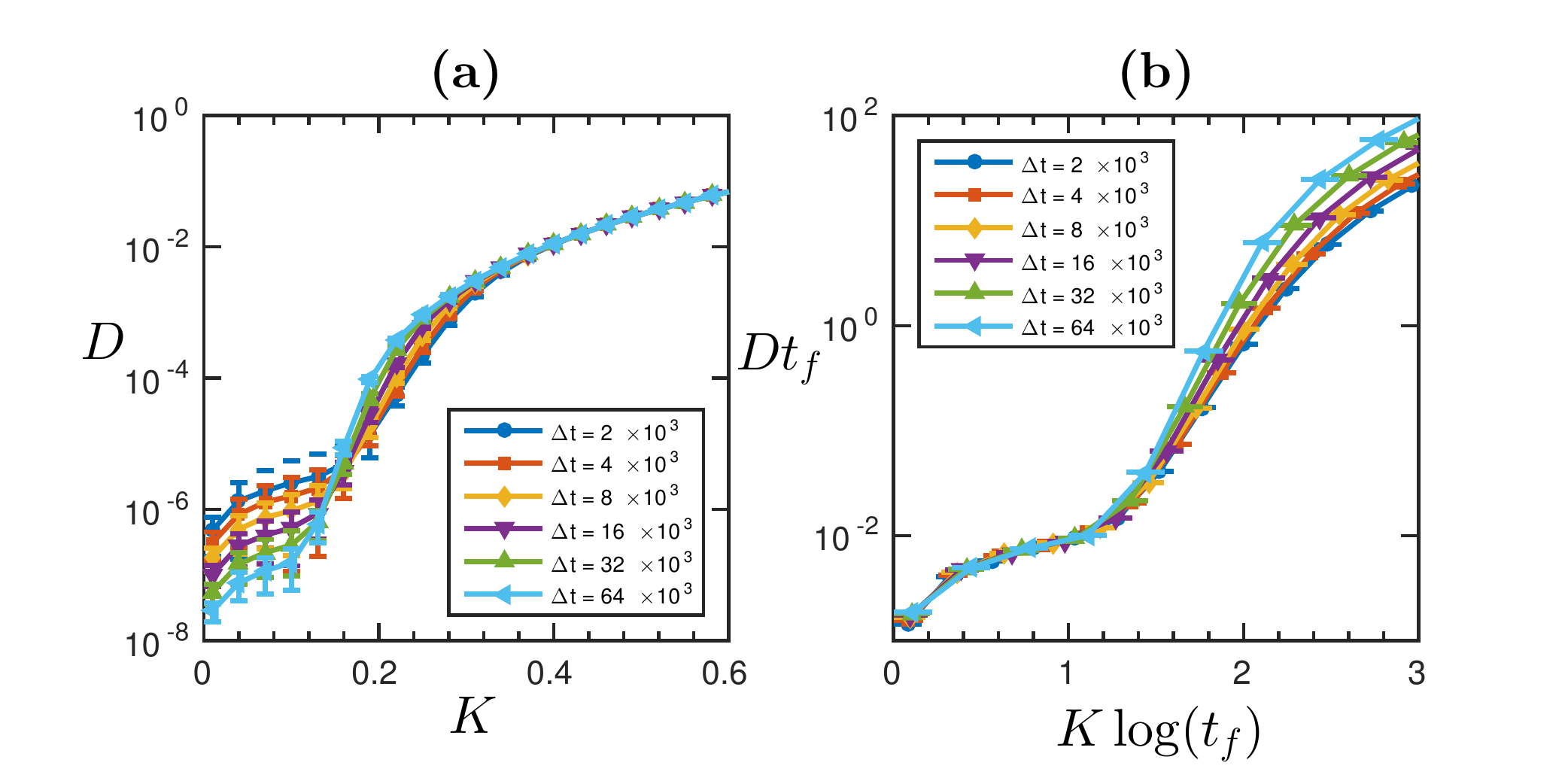}
\includegraphics[trim={1.2cm 0.5cm 1.8cm 0.7cm},clip,scale=0.46]{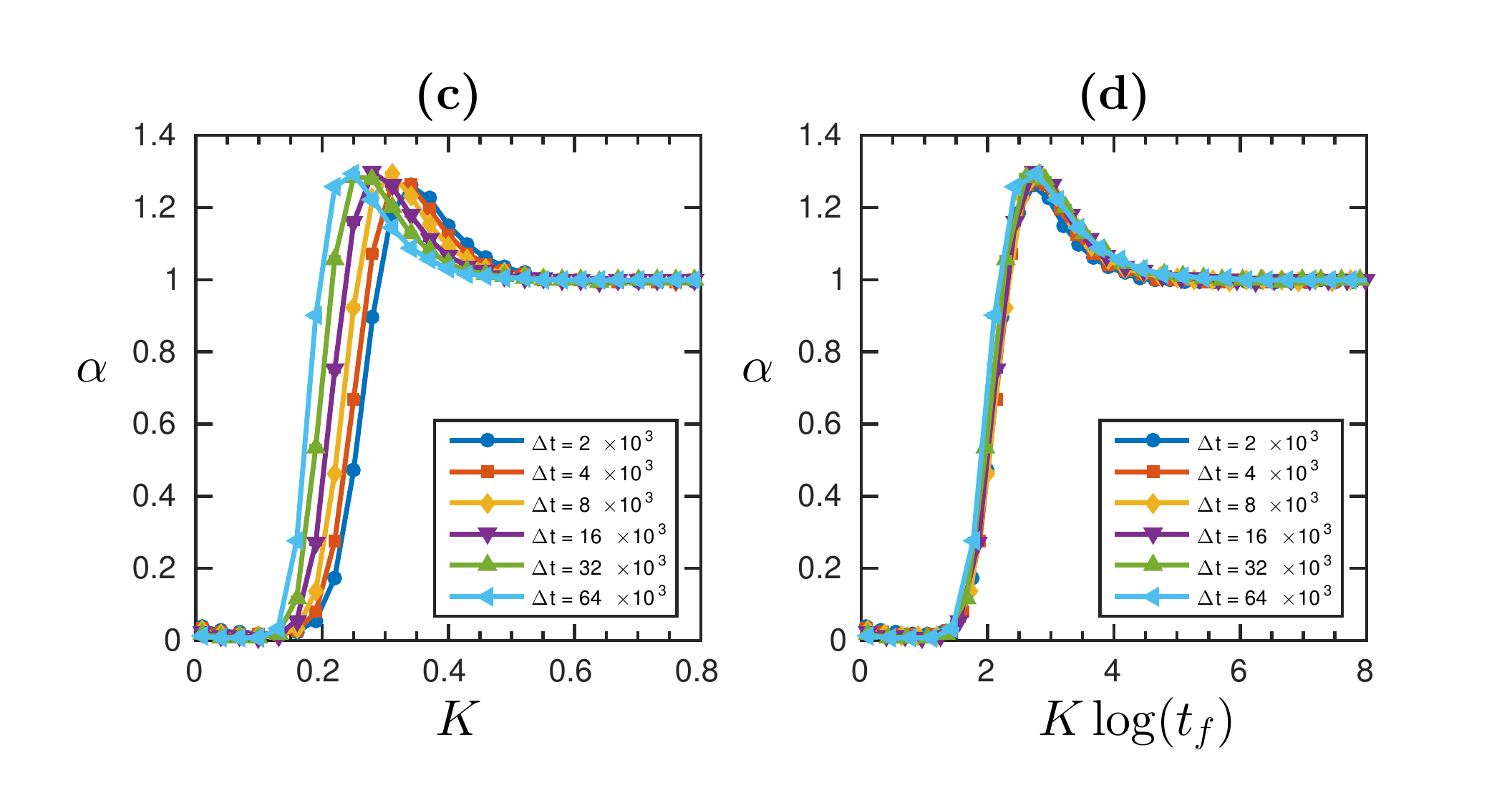}
\caption{(a) Diffusion coefficient $D$ and (b) their collapse for $K\lesssim 1.5/\log(t_f)$ with 
statistical error in the data points for the same initial conditions as in Fig.~\ref{dc_posK}.
(c) and (d) are the same plots as in Fig.~\ref{exp_posK} with statistical error in the data 
points indicated by the symbol sizes.}
\label{d_alpha_error}
\end{figure}

We now consider the scenario of marginal localization as shown in Figs.~\ref{dc_posK} 
and \ref{exp_posK} as an example to show statistical error in our data due to random initial 
conditions. Fig.~\ref{d_alpha_error} shows the  statistical error in diffusion 
constant $D$, determined by  Eq.~(\ref{dc_def}) with $1000$ random initial conditions. 
On the other hand, the error in $\alpha$ is given by the error in fitting $\langle p^2(t)\rangle$ 
as a general power-law function $At^{\alpha}$.

\vskip 0.5cm
\noindent{\bf References}
\vskip 0.5cm

\bibliographystyle{naturemag}
\bibliography{ref}

\begin{thebibliography}{10}
\expandafter\ifx\csname url\endcsname\relax
  \def\url#1{\texttt{#1}}\fi
\expandafter\ifx\csname urlprefix\endcsname\relax\def\urlprefix{URL }\fi
\providecommand{\bibinfo}[2]{#2}
\providecommand{\eprint}[2][]{\url{#2}}

\bibitem{eckardt17atomic}
\bibinfo{author}{Eckardt, A.}
\newblock \bibinfo{title}{Atomic quantum gases in periodically driven optical
  lattices}.
\newblock \emph{\bibinfo{journal}{Review of Modern Physics}}
  \textbf{\bibinfo{volume}{89}}, \bibinfo{pages}{011004}
  (\bibinfo{year}{2017}).

\bibitem{fausti11light}
\bibinfo{author}{Fausti, D.} \emph{et~al.}
\newblock \bibinfo{title}{Light-induced superconductivity in a stripe-ordered
  cuprate}.
\newblock \emph{\bibinfo{journal}{science}} \textbf{\bibinfo{volume}{331}},
  \bibinfo{pages}{189--191} (\bibinfo{year}{2011}).

\bibitem{hu13optically}
\bibinfo{author}{Hu, W.} \emph{et~al.}
\newblock \bibinfo{title}{Optically enhanced coherent transport in yba2cu3o6.5
  by ultrafast redistribution of interlayer coupling}.
\newblock \emph{\bibinfo{journal}{Nature Materials}}
  \textbf{\bibinfo{volume}{13}}, \bibinfo{pages}{705--711}
  (\bibinfo{year}{2014}).

\bibitem{babadi2017theory}
\bibinfo{author}{Babadi, M.}, \bibinfo{author}{Knap, M.},
  \bibinfo{author}{Martin, I.}, \bibinfo{author}{Refael, G.} \&
  \bibinfo{author}{Demler, E.}
\newblock \bibinfo{title}{The theory of parametrically amplified
  electron-phonon superconductivity}.
\newblock \emph{\bibinfo{journal}{arXiv preprint arXiv:1702.02531}}
  (\bibinfo{year}{2017}).

\bibitem{sentef2017light}
\bibinfo{author}{Sentef, M.}
\newblock \bibinfo{title}{Light-enhanced electron-phonon coupling from
  nonlinear electron-phonon coupling}.
\newblock \emph{\bibinfo{journal}{Physical Review B}}
  \textbf{\bibinfo{volume}{95}}, \bibinfo{pages}{205111}
  (\bibinfo{year}{2017}).

\bibitem{kennes2016electronic}
\bibinfo{author}{Kennes, D.~M.}, \bibinfo{author}{Wilner, E.~Y.},
  \bibinfo{author}{Reichman, D.~R.} \& \bibinfo{author}{Millis, A.~J.}
\newblock \bibinfo{title}{Electronic squeezing of pumped phonons: Negative u
  and transient superconductivity}.
\newblock \emph{\bibinfo{journal}{arXiv preprint arXiv:1609.03802}}
  (\bibinfo{year}{2016}).

\bibitem{murakami2017nonequilibrium}
\bibinfo{author}{Murakami, Y.}, \bibinfo{author}{Tsuji, N.},
  \bibinfo{author}{Eckstein, M.} \& \bibinfo{author}{Werner, P.}
\newblock \bibinfo{title}{Nonequilibrium steady states and transient dynamics
  of superconductors under phonon driving}.
\newblock \emph{\bibinfo{journal}{arXiv preprint arXiv:1702.02942}}
  (\bibinfo{year}{2017}).

\bibitem{oka09photovoltaic}
\bibinfo{author}{Oka, T.} \& \bibinfo{author}{Aoki, H.}
\newblock \bibinfo{title}{Photovoltaic hall effect in graphene}.
\newblock \emph{\bibinfo{journal}{Physical Review B}}
  \textbf{\bibinfo{volume}{79}}, \bibinfo{pages}{081406}
  (\bibinfo{year}{2009}).

\bibitem{lindner11floquet}
\bibinfo{author}{Lindner, N.~H.}, \bibinfo{author}{Refael, G.} \&
  \bibinfo{author}{Galitski, V.}
\newblock \bibinfo{title}{Floquet topological insulator in semiconductor
  quantum wells}.
\newblock \emph{\bibinfo{journal}{Nature Physics}}
  \textbf{\bibinfo{volume}{7}}, \bibinfo{pages}{490--495}
  (\bibinfo{year}{2011}).

\bibitem{kitagawa11transport}
\bibinfo{author}{Kitagawa, T.}, \bibinfo{author}{Oka, T.},
  \bibinfo{author}{Brataas, A.}, \bibinfo{author}{Fu, L.} \&
  \bibinfo{author}{Demler, E.}
\newblock \bibinfo{title}{Transport properties of nonequilibrium systems under
  the application of light: Photoinduced quantum hall insulators without landau
  levels}.
\newblock \emph{\bibinfo{journal}{Physical Review B}}
  \textbf{\bibinfo{volume}{84}}, \bibinfo{pages}{235108}
  (\bibinfo{year}{2011}).

\bibitem{rechtsman13photonic}
\bibinfo{author}{Rechtsman, M.~C.} \emph{et~al.}
\newblock \bibinfo{title}{Photonic floquet topological insulators}.
\newblock \emph{\bibinfo{journal}{Nature}} \textbf{\bibinfo{volume}{496}},
  \bibinfo{pages}{196--200} (\bibinfo{year}{2013}).

\bibitem{rudner13anomalous}
\bibinfo{author}{Rudner, M.~S.}, \bibinfo{author}{Lindner, N.~H.},
  \bibinfo{author}{Berg, E.} \& \bibinfo{author}{Levin, M.}
\newblock \bibinfo{title}{Anomalous edge states and the bulk-edge
  correspondence for periodically driven two-dimensional systems}.
\newblock \emph{\bibinfo{journal}{Physical Review X}}
  \textbf{\bibinfo{volume}{3}}, \bibinfo{pages}{031005} (\bibinfo{year}{2013}).

\bibitem{jotzu14experimental}
\bibinfo{author}{Jotzu, G.} \emph{et~al.}
\newblock \bibinfo{title}{Experimental realization of the topological haldane
  model with ultracold fermions}.
\newblock \emph{\bibinfo{journal}{Nature}} \textbf{\bibinfo{volume}{515}},
  \bibinfo{pages}{237--240} (\bibinfo{year}{2014}).

\bibitem{fleury16floquet}
\bibinfo{author}{Fleury, R.}, \bibinfo{author}{Khanikaev, A.~B.} \&
  \bibinfo{author}{Alu, A.}
\newblock \bibinfo{title}{Floquet topological insulators for sound}.
\newblock \emph{\bibinfo{journal}{Nature communications}}
  \textbf{\bibinfo{volume}{7}}, \bibinfo{pages}{11744} (\bibinfo{year}{2016}).

\bibitem{khemani16phase}
\bibinfo{author}{Khemani, V.}, \bibinfo{author}{Lazarides, A.},
  \bibinfo{author}{Moessner, R.} \& \bibinfo{author}{Sondhi, S.~L.}
\newblock \bibinfo{title}{Phase structure of driven quantum systems}.
\newblock \emph{\bibinfo{journal}{Physical Review Letters}}
  \textbf{\bibinfo{volume}{116}}, \bibinfo{pages}{250401}
  (\bibinfo{year}{2016}).

\bibitem{else16floquet}
\bibinfo{author}{Else, D.~V.}, \bibinfo{author}{Bauer, B.} \&
  \bibinfo{author}{Nayak, C.}
\newblock \bibinfo{title}{Floquet time crystals}.
\newblock \emph{\bibinfo{journal}{Physical Review Letters}}
  \textbf{\bibinfo{volume}{117}}, \bibinfo{pages}{090402}
  (\bibinfo{year}{2016}).

\bibitem{zhang17observation}
\bibinfo{author}{Zhang, J.} \emph{et~al.}
\newblock \bibinfo{title}{Observation of a discrete time crystal}.
\newblock \emph{\bibinfo{journal}{Nature}} \textbf{\bibinfo{volume}{543}},
  \bibinfo{pages}{217--220} (\bibinfo{year}{2017}).

\bibitem{moessner2017equilibration}
\bibinfo{author}{Moessner, R.} \& \bibinfo{author}{Sondhi, S.}
\newblock \bibinfo{title}{Equilibration and order in quantum floquet matter}.
\newblock \emph{\bibinfo{journal}{Nature Physics}}
  \textbf{\bibinfo{volume}{13}}, \bibinfo{pages}{424} (\bibinfo{year}{2017}).

\bibitem{russomanno2017spin}
\bibinfo{author}{Russomanno, A.}, \bibinfo{author}{Dalla~Torre, E.~G.}
  \emph{et~al.}
\newblock \bibinfo{title}{Spin and topological order in a periodically driven
  spin chain}.
\newblock \emph{\bibinfo{journal}{Physical Review B}}
  \textbf{\bibinfo{volume}{96}}, \bibinfo{pages}{045422}
  (\bibinfo{year}{2017}).

\bibitem{yao17discrete}
\bibinfo{author}{Yao, N.~Y.}, \bibinfo{author}{Potter, A.~C.},
  \bibinfo{author}{Potirniche, I.-D.} \& \bibinfo{author}{Vishwanath, A.}
\newblock \bibinfo{title}{Discrete time crystals: rigidity, criticality, and
  realizations}.
\newblock \emph{\bibinfo{journal}{Physical Review Letters}}
  \textbf{\bibinfo{volume}{118}}, \bibinfo{pages}{030401}
  (\bibinfo{year}{2017}).

\bibitem{mori2017thermalization}
\bibinfo{author}{{Mori}, T.}, \bibinfo{author}{{Ikeda}, T.~N.},
  \bibinfo{author}{{Kaminishi}, E.} \& \bibinfo{author}{{Ueda}, M.}
\newblock \bibinfo{title}{{Thermalization and prethermalization in isolated
  quantum systems: a theoretical overview}}.
\newblock \emph{\bibinfo{journal}{ArXiv e-prints}}  (\bibinfo{year}{2017}).
\newblock \eprint{1712.08790}.

\bibitem{d13many}
\bibinfo{author}{D’Alessio, L.} \& \bibinfo{author}{Polkovnikov, A.}
\newblock \bibinfo{title}{Many-body energy localization transition in
  periodically driven systems}.
\newblock \emph{\bibinfo{journal}{Annals of Physics}}
  \textbf{\bibinfo{volume}{333}}, \bibinfo{pages}{19--33}
  (\bibinfo{year}{2013}).

\bibitem{d14long}
\bibinfo{author}{D’Alessio, L.} \& \bibinfo{author}{Rigol, M.}
\newblock \bibinfo{title}{Long-time behavior of isolated periodically driven
  interacting lattice systems}.
\newblock \emph{\bibinfo{journal}{Physical Review X}}
  \textbf{\bibinfo{volume}{4}}, \bibinfo{pages}{041048} (\bibinfo{year}{2014}).

\bibitem{lazarides14equilibrium}
\bibinfo{author}{Lazarides, A.}, \bibinfo{author}{Das, A.} \&
  \bibinfo{author}{Moessner, R.}
\newblock \bibinfo{title}{Equilibrium states of generic quantum systems subject
  to periodic driving}.
\newblock \emph{\bibinfo{journal}{Physical Review E}}
  \textbf{\bibinfo{volume}{90}}, \bibinfo{pages}{012110}
  (\bibinfo{year}{2014}).

\bibitem{russomanno15asymptotic}
\bibinfo{author}{Russomanno, A.}, \bibinfo{author}{Sharma, S.},
  \bibinfo{author}{Dutta, A.} \& \bibinfo{author}{Santoro, G.~E.}
\newblock \bibinfo{title}{Asymptotic work statistics of periodically driven
  ising chains}.
\newblock \emph{\bibinfo{journal}{Journal of Statistical Mechanics: Theory and
  Experiment}} \textbf{\bibinfo{volume}{2015}}, \bibinfo{pages}{P08030}
  (\bibinfo{year}{2015}).

\bibitem{russomanno15thermalization}
\bibinfo{author}{Russomanno, A.}, \bibinfo{author}{Fazio, R.} \&
  \bibinfo{author}{Santoro, G.~E.}
\newblock \bibinfo{title}{Thermalization in a periodically driven fully
  connected quantum ising ferromagnet}.
\newblock \emph{\bibinfo{journal}{EPL (Europhysics Letters)}}
  \textbf{\bibinfo{volume}{110}}, \bibinfo{pages}{37005}
  (\bibinfo{year}{2015}).

\bibitem{seetharam15controlled}
\bibinfo{author}{Seetharam, K.~I.}, \bibinfo{author}{Bardyn, C.-E.},
  \bibinfo{author}{Lindner, N.~H.}, \bibinfo{author}{Rudner, M.~S.} \&
  \bibinfo{author}{Refael, G.}
\newblock \bibinfo{title}{Controlled population of floquet-bloch states via
  coupling to bose and fermi baths}.
\newblock \emph{\bibinfo{journal}{Physical Review X}}
  \textbf{\bibinfo{volume}{5}}, \bibinfo{pages}{041050} (\bibinfo{year}{2015}).

\bibitem{prosen98time}
\bibinfo{author}{Prosen, T.}
\newblock \bibinfo{title}{Time evolution of a quantum many-body system:
  Transition from integrability to ergodicity in the thermodynamic limit}.
\newblock \emph{\bibinfo{journal}{Physical Review Letters}}
  \textbf{\bibinfo{volume}{80}}, \bibinfo{pages}{1808} (\bibinfo{year}{1998}).

\bibitem{citro15dynamical}
\bibinfo{author}{Citro, R.} \emph{et~al.}
\newblock \bibinfo{title}{Dynamical stability of a many-body kapitza pendulum}.
\newblock \emph{\bibinfo{journal}{Annals of Physics}}
  \textbf{\bibinfo{volume}{360}}, \bibinfo{pages}{694--710}
  (\bibinfo{year}{2015}).

\bibitem{bhadra2018dynamics}
\bibinfo{author}{Bhadra, N.}
\newblock \bibinfo{title}{Dynamics of a system of coupled inverted pendula with
  vertical forcing}.
\newblock \emph{\bibinfo{journal}{arXiv preprint arXiv:1803.01643}}
  (\bibinfo{year}{2018}).

\bibitem{russomanno12periodic}
\bibinfo{author}{Russomanno, A.}, \bibinfo{author}{Silva, A.} \&
  \bibinfo{author}{Santoro, G.~E.}
\newblock \bibinfo{title}{Periodic steady regime and interference in a
  periodically driven quantum system}.
\newblock \emph{\bibinfo{journal}{Physical Review Letters}}
  \textbf{\bibinfo{volume}{109}}, \bibinfo{pages}{257201}
  (\bibinfo{year}{2012}).

\bibitem{russomanno2016kibble}
\bibinfo{author}{Russomanno, A.} \& \bibinfo{author}{Dalla~Torre, E.~G.}
\newblock \bibinfo{title}{Kibble-zurek scaling in periodically driven quantum
  systems}.
\newblock \emph{\bibinfo{journal}{EPL (Europhysics Letters)}}
  \textbf{\bibinfo{volume}{115}}, \bibinfo{pages}{30006}
  (\bibinfo{year}{2016}).

\bibitem{ponte15many}
\bibinfo{author}{Ponte, P.}, \bibinfo{author}{Papi{\'c}, Z.},
  \bibinfo{author}{Huveneers, F.} \& \bibinfo{author}{Abanin, D.~A.}
\newblock \bibinfo{title}{Many-body localization in periodically driven
  systems}.
\newblock \emph{\bibinfo{journal}{Physical Review Letters}}
  \textbf{\bibinfo{volume}{114}}, \bibinfo{pages}{140401}
  (\bibinfo{year}{2015}).

\bibitem{lazarides15fate}
\bibinfo{author}{Lazarides, A.}, \bibinfo{author}{Das, A.} \&
  \bibinfo{author}{Moessner, R.}
\newblock \bibinfo{title}{Fate of many-body localization under periodic
  driving}.
\newblock \emph{\bibinfo{journal}{Physical Review Letters}}
  \textbf{\bibinfo{volume}{115}}, \bibinfo{pages}{030402}
  (\bibinfo{year}{2015}).

\bibitem{ponte15periodically}
\bibinfo{author}{Ponte, P.}, \bibinfo{author}{Chandran, A.},
  \bibinfo{author}{Papi{\'c}, Z.} \& \bibinfo{author}{Abanin, D.~A.}
\newblock \bibinfo{title}{Periodically driven ergodic and many-body localized
  quantum systems}.
\newblock \emph{\bibinfo{journal}{Annals of Physics}}
  \textbf{\bibinfo{volume}{353}}, \bibinfo{pages}{196--204}
  (\bibinfo{year}{2015}).

\bibitem{abanin16theory}
\bibinfo{author}{Abanin, D.~A.}, \bibinfo{author}{De~Roeck, W.} \&
  \bibinfo{author}{Huveneers, F.}
\newblock \bibinfo{title}{Theory of many-body localization in periodically
  driven systems}.
\newblock \emph{\bibinfo{journal}{Annals of Physics}}
  \textbf{\bibinfo{volume}{372}}, \bibinfo{pages}{1--11}
  (\bibinfo{year}{2016}).

\bibitem{agarwal17localization}
\bibinfo{author}{Agarwal, K.}, \bibinfo{author}{Ganeshan, S.} \&
  \bibinfo{author}{Bhatt, R.}
\newblock \bibinfo{title}{Localization and transport in a strongly driven
  anderson insulator}.
\newblock \emph{\bibinfo{journal}{Physical Review B}}
  \textbf{\bibinfo{volume}{96}}, \bibinfo{pages}{014201}
  (\bibinfo{year}{2017}).

\bibitem{dumitrescu17logarithmically}
\bibinfo{author}{Dumitrescu, P.~T.}, \bibinfo{author}{Vasseur, R.} \&
  \bibinfo{author}{Potter, A.~C.}
\newblock \bibinfo{title}{Logarithmically slow relaxation in quasi-periodically
  driven random spin chains}.
\newblock \emph{\bibinfo{journal}{arXiv preprint arXiv:1708.00865}}
  (\bibinfo{year}{2017}).

\bibitem{choudhury14stability}
\bibinfo{author}{Choudhury, S.} \& \bibinfo{author}{Mueller, E.~J.}
\newblock \bibinfo{title}{Stability of a floquet bose-einstein condensate in a
  one-dimensional optical lattice}.
\newblock \emph{\bibinfo{journal}{Physical Review A}}
  \textbf{\bibinfo{volume}{90}}, \bibinfo{pages}{013621}
  (\bibinfo{year}{2014}).

\bibitem{bukov15prethermal}
\bibinfo{author}{Bukov, M.}, \bibinfo{author}{Gopalakrishnan, S.},
  \bibinfo{author}{Knap, M.} \& \bibinfo{author}{Demler, E.}
\newblock \bibinfo{title}{Prethermal floquet steady states and instabilities in
  the periodically driven, weakly interacting bose-hubbard model}.
\newblock \emph{\bibinfo{journal}{Physical Review Letters}}
  \textbf{\bibinfo{volume}{115}}, \bibinfo{pages}{205301}
  (\bibinfo{year}{2015}).

\bibitem{abanin15exponentially}
\bibinfo{author}{Abanin, D.~A.}, \bibinfo{author}{De~Roeck, W.} \&
  \bibinfo{author}{Huveneers, F.}
\newblock \bibinfo{title}{Exponentially slow heating in periodically driven
  many-body systems}.
\newblock \emph{\bibinfo{journal}{Physical Review Letters}}
  \textbf{\bibinfo{volume}{115}}, \bibinfo{pages}{256803}
  (\bibinfo{year}{2015}).

\bibitem{goldman15periodically}
\bibinfo{author}{Goldman, N.}, \bibinfo{author}{Dalibard, J.},
  \bibinfo{author}{Aidelsburger, M.} \& \bibinfo{author}{Cooper, N.}
\newblock \bibinfo{title}{Periodically driven quantum matter: The case of
  resonant modulations}.
\newblock \emph{\bibinfo{journal}{Physical Review A}}
  \textbf{\bibinfo{volume}{91}}, \bibinfo{pages}{033632}
  (\bibinfo{year}{2015}).

\bibitem{chandran16interaction}
\bibinfo{author}{Chandran, A.} \& \bibinfo{author}{Sondhi, S.~L.}
\newblock \bibinfo{title}{Interaction-stabilized steady states in the driven o
  (n) model}.
\newblock \emph{\bibinfo{journal}{Physical Review B}}
  \textbf{\bibinfo{volume}{93}}, \bibinfo{pages}{174305}
  (\bibinfo{year}{2016}).

\bibitem{mori2016rigorous}
\bibinfo{author}{Mori, T.}, \bibinfo{author}{Kuwahara, T.} \&
  \bibinfo{author}{Saito, K.}
\newblock \bibinfo{title}{Rigorous bound on energy absorption and generic
  relaxation in periodically driven quantum systems}.
\newblock \emph{\bibinfo{journal}{Physical review letters}}
  \textbf{\bibinfo{volume}{116}}, \bibinfo{pages}{120401}
  (\bibinfo{year}{2016}).

\bibitem{abanin17rigorous}
\bibinfo{author}{Abanin, D.}, \bibinfo{author}{De~Roeck, W.},
  \bibinfo{author}{Ho, W.~W.} \& \bibinfo{author}{Huveneers, F.}
\newblock \bibinfo{title}{A rigorous theory of many-body prethermalization for
  periodically driven and closed quantum systems}.
\newblock \emph{\bibinfo{journal}{Communications in Mathematical Physics}}
  \textbf{\bibinfo{volume}{354}}, \bibinfo{pages}{809--827}
  (\bibinfo{year}{2017}).

\bibitem{lellouch17parametric}
\bibinfo{author}{Lellouch, S.}, \bibinfo{author}{Bukov, M.},
  \bibinfo{author}{Demler, E.} \& \bibinfo{author}{Goldman, N.}
\newblock \bibinfo{title}{Parametric instability rates in periodically driven
  band systems}.
\newblock \emph{\bibinfo{journal}{Physical Review X}}
  \textbf{\bibinfo{volume}{7}}, \bibinfo{pages}{021015} (\bibinfo{year}{2017}).

\bibitem{lellouch17parametric1}
\bibinfo{author}{Lellouch, S.} \& \bibinfo{author}{Goldman, N.}
\newblock \bibinfo{title}{Parametric instabilities in resonantly-driven
  bose-einstein condensates}.
\newblock \emph{\bibinfo{journal}{arXiv preprint arXiv:1711.08832}}
  (\bibinfo{year}{2017}).

\bibitem{machado17exponentially}
\bibinfo{author}{Machado, F.}, \bibinfo{author}{Meyer, G.~D.},
  \bibinfo{author}{Else, D.~V.}, \bibinfo{author}{Nayak, C.} \&
  \bibinfo{author}{Yao, N.~Y.}
\newblock \bibinfo{title}{Exponentially slow heating in short and long-range
  interacting floquet systems}.
\newblock \emph{\bibinfo{journal}{arXiv preprint arXiv:1708.01620}}
  (\bibinfo{year}{2017}).

\bibitem{kuwahara16floquet}
\bibinfo{author}{Kuwahara, T.}, \bibinfo{author}{Mori, T.} \&
  \bibinfo{author}{Saito, K.}
\newblock \bibinfo{title}{Floquet--magnus theory and generic transient dynamics
  in periodically driven many-body quantum systems}.
\newblock \emph{\bibinfo{journal}{Annals of Physics}}
  \textbf{\bibinfo{volume}{367}}, \bibinfo{pages}{96--124}
  (\bibinfo{year}{2016}).

\bibitem{canovi16stroboscopic}
\bibinfo{author}{Canovi, E.}, \bibinfo{author}{Kollar, M.} \&
  \bibinfo{author}{Eckstein, M.}
\newblock \bibinfo{title}{Stroboscopic prethermalization in weakly interacting
  periodically driven systems}.
\newblock \emph{\bibinfo{journal}{Physical Review E}}
  \textbf{\bibinfo{volume}{93}}, \bibinfo{pages}{012130}
  (\bibinfo{year}{2016}).

\bibitem{weidinger17floquet}
\bibinfo{author}{Weidinger, S.~A.} \& \bibinfo{author}{Knap, M.}
\newblock \bibinfo{title}{Floquet prethermalization and regimes of heating in a
  periodically driven, interacting quantum system}.
\newblock \emph{\bibinfo{journal}{Scientific Reports}}
  \textbf{\bibinfo{volume}{7}}, \bibinfo{pages}{45382} (\bibinfo{year}{2017}).

\bibitem{zeng17prethermal}
\bibinfo{author}{Zeng, T.-S.} \& \bibinfo{author}{Sheng, D.}
\newblock \bibinfo{title}{Prethermal time crystals in a one-dimensional
  periodically driven floquet system}.
\newblock \emph{\bibinfo{journal}{Physical Review B}}
  \textbf{\bibinfo{volume}{96}}, \bibinfo{pages}{094202}
  (\bibinfo{year}{2017}).

\bibitem{else17prethermal}
\bibinfo{author}{Else, D.~V.}, \bibinfo{author}{Bauer, B.} \&
  \bibinfo{author}{Nayak, C.}
\newblock \bibinfo{title}{Prethermal phases of matter protected by
  time-translation symmetry}.
\newblock \emph{\bibinfo{journal}{Physical Review X}}
  \textbf{\bibinfo{volume}{7}}, \bibinfo{pages}{011026} (\bibinfo{year}{2017}).

\bibitem{abanin17effective}
\bibinfo{author}{Abanin, D.~A.}, \bibinfo{author}{De~Roeck, W.},
  \bibinfo{author}{Ho, W.~W.} \& \bibinfo{author}{Huveneers, F.}
\newblock \bibinfo{title}{Effective hamiltonians, prethermalization, and slow
  energy absorption in periodically driven many-body systems}.
\newblock \emph{\bibinfo{journal}{Physical Review B}}
  \textbf{\bibinfo{volume}{95}}, \bibinfo{pages}{014112}
  (\bibinfo{year}{2017}).

\bibitem{chirikov79a}
\bibinfo{author}{Chirikov, B.~V.}
\newblock \bibinfo{title}{A universal instability of many-dimensional
  oscillator systems}.
\newblock \emph{\bibinfo{journal}{Physics Reports}}
  \textbf{\bibinfo{volume}{52}}, \bibinfo{pages}{263} (\bibinfo{year}{1979}).

\bibitem{kaneko89diffusion}
\bibinfo{author}{Kaneko, K.} \& \bibinfo{author}{Konishi, T.}
\newblock \bibinfo{title}{Diffusion in hamiltonian dynamical systems with many
  degrees of freedom}.
\newblock \emph{\bibinfo{journal}{Physical Review A}}
  \textbf{\bibinfo{volume}{40}}, \bibinfo{pages}{6130} (\bibinfo{year}{1989}).

\bibitem{konishi90diffusion}
\bibinfo{author}{Konishi, T.} \& \bibinfo{author}{Kaneko, K.}
\newblock \bibinfo{title}{Diffusion in hamiltonian chaos and its size
  dependence}.
\newblock \emph{\bibinfo{journal}{Journal of Physics A: Mathematical and
  General}} \textbf{\bibinfo{volume}{23}}, \bibinfo{pages}{L715}
  (\bibinfo{year}{1990}).

\bibitem{weigert02quantum}
\bibinfo{author}{Weigert, S.}
\newblock \bibinfo{title}{Quantum parametric resonance}.
\newblock \emph{\bibinfo{journal}{Journal of Physics A: Mathematical and
  General}} \textbf{\bibinfo{volume}{35}}, \bibinfo{pages}{4169}
  (\bibinfo{year}{2002}).

\bibitem{greene79method}
\bibinfo{author}{Greene, J.~M.}
\newblock \bibinfo{title}{A method for determining a stochastic transition}.
\newblock \emph{\bibinfo{journal}{Journal of Mathematical Physics}}
  \textbf{\bibinfo{volume}{20}}, \bibinfo{pages}{1183--1201}
  (\bibinfo{year}{1979}).

\bibitem{arnold68}
\bibinfo{author}{Arnold, V.~I.} \& \bibinfo{author}{Avez, A.}
\newblock \emph{\bibinfo{title}{{Ergodic Problems in Classical Mechanics}}}
  (\bibinfo{publisher}{Benjamin Cummings, Reading, MA}, \bibinfo{year}{1968}).

\bibitem{fishman82chaos}
\bibinfo{author}{Fishman, S.}, \bibinfo{author}{Grempel, D.} \&
  \bibinfo{author}{Prange, R.}
\newblock \bibinfo{title}{Chaos, quantum recurrences, and anderson
  localization}.
\newblock \emph{\bibinfo{journal}{Physical Review Letters}}
  \textbf{\bibinfo{volume}{49}}, \bibinfo{pages}{509} (\bibinfo{year}{1982}).

\bibitem{fishman89scaling}
\bibinfo{author}{Fishman, S.}, \bibinfo{author}{Prange, R.} \&
  \bibinfo{author}{Griniasty, M.}
\newblock \bibinfo{title}{Scaling theory for the localization length of the
  kicked rotor}.
\newblock \emph{\bibinfo{journal}{Physical Review A}}
  \textbf{\bibinfo{volume}{39}}, \bibinfo{pages}{1628} (\bibinfo{year}{1989}).

\bibitem{anderson58absence}
\bibinfo{author}{Anderson, P.~W.}
\newblock \bibinfo{title}{Absence of diffusion in certain random lattices}.
\newblock \emph{\bibinfo{journal}{Physical review}}
  \textbf{\bibinfo{volume}{109}}, \bibinfo{pages}{1492} (\bibinfo{year}{1958}).

\bibitem{notarnicola17from}
\bibinfo{author}{Notarnicola, S.} \emph{et~al.}
\newblock \bibinfo{title}{From localization to anomalous diffusion in the
  dynamics of coupled kicked rotors}.
\newblock \emph{\bibinfo{journal}{arXiv preprint arXiv:1709.05657}}
  (\bibinfo{year}{2017}).

\bibitem{seetharam17absence}
\bibinfo{author}{Seetharam, K.~I.}, \bibinfo{author}{Titum, P.},
  \bibinfo{author}{Kolodrubetz, M.} \& \bibinfo{author}{Refael, G.}
\newblock \bibinfo{title}{Absence of thermalization in finite isolated
  interacting floquet systems}.
\newblock \emph{\bibinfo{journal}{arXiv preprint arXiv:1710.09843}}
  (\bibinfo{year}{2017}).

\bibitem{keser16dynamical}
\bibinfo{author}{Keser, A.~C.}, \bibinfo{author}{Ganeshan, S.},
  \bibinfo{author}{Refael, G.} \& \bibinfo{author}{Galitski, V.}
\newblock \bibinfo{title}{Dynamical many-body localization in an integrable
  model}.
\newblock \emph{\bibinfo{journal}{Physical Review B}}
  \textbf{\bibinfo{volume}{94}}, \bibinfo{pages}{085120}
  (\bibinfo{year}{2016}).

\bibitem{cryan09field}
\bibinfo{author}{Cryan, J.~P.}, \bibinfo{author}{Bucksbaum, P.~H.} \&
  \bibinfo{author}{Coffee, R.~N.}
\newblock \bibinfo{title}{Field-free alignment in repetitively kicked nitrogen
  gas}.
\newblock \emph{\bibinfo{journal}{Physical Review A}}
  \textbf{\bibinfo{volume}{80}}, \bibinfo{pages}{063412}
  (\bibinfo{year}{2009}).

\bibitem{zhdanovich12quantum}
\bibinfo{author}{Zhdanovich, S.} \emph{et~al.}
\newblock \bibinfo{title}{Quantum resonances in selective rotational excitation
  of molecules with a sequence of ultrashort laser pulses}.
\newblock \emph{\bibinfo{journal}{Physical Review Letters}}
  \textbf{\bibinfo{volume}{109}}, \bibinfo{pages}{043003}
  (\bibinfo{year}{2012}).

\bibitem{zahedpour14quantum}
\bibinfo{author}{Zahedpour, S.}, \bibinfo{author}{Wahlstrand, J.} \&
  \bibinfo{author}{Milchberg, H.}
\newblock \bibinfo{title}{Quantum control of molecular gas hydrodynamics}.
\newblock \emph{\bibinfo{journal}{Physical Review Letters}}
  \textbf{\bibinfo{volume}{112}}, \bibinfo{pages}{143601}
  (\bibinfo{year}{2014}).

\bibitem{gadway13evidence}
\bibinfo{author}{Gadway, B.}, \bibinfo{author}{Reeves, J.},
  \bibinfo{author}{Krinner, L.} \& \bibinfo{author}{Schneble, D.}
\newblock \bibinfo{title}{Evidence for a quantum-to-classical transition in a
  pair of coupled quantum rotors}.
\newblock \emph{\bibinfo{journal}{Physical Review Letters}}
  \textbf{\bibinfo{volume}{110}}, \bibinfo{pages}{190401}
  (\bibinfo{year}{2013}).

\bibitem{chirikov97arnold}
\bibinfo{author}{Chirikov, B.} \& \bibinfo{author}{Vecheslavov, V.}
\newblock \bibinfo{title}{Arnold diffusion in large systems}.
\newblock \emph{\bibinfo{journal}{Journal of Experimental and Theoretical
  Physics}} \textbf{\bibinfo{volume}{85}}, \bibinfo{pages}{616--624}
  (\bibinfo{year}{1997}).

\bibitem{falcioni91ergodic}
\bibinfo{author}{Falcioni, M.}, \bibinfo{author}{Marconi, U. M.~B.} \&
  \bibinfo{author}{Vulpiani, A.}
\newblock \bibinfo{title}{Ergodic properties of high-dimensional symplectic
  maps}.
\newblock \emph{\bibinfo{journal}{Physical Review A}}
  \textbf{\bibinfo{volume}{44}}, \bibinfo{pages}{2263} (\bibinfo{year}{1991}).

\bibitem{mulansky11strong}
\bibinfo{author}{Mulansky, M.}, \bibinfo{author}{Ahnert, K.},
  \bibinfo{author}{Pikovsky, A.} \& \bibinfo{author}{Shepelyansky, D.~L.}
\newblock \bibinfo{title}{Strong and weak chaos in weakly nonintegrable
  many-body hamiltonian systems}.
\newblock \emph{\bibinfo{journal}{Journal of Statistical Physics}}
  \textbf{\bibinfo{volume}{145}}, \bibinfo{pages}{1256--1274}
  (\bibinfo{year}{2011}).

\bibitem{chirikov79universal}
\bibinfo{author}{Chirikov, B.~V.}
\newblock \bibinfo{title}{A universal instability of many-dimensional
  oscillator systems}.
\newblock \emph{\bibinfo{journal}{Physics reports}}
  \textbf{\bibinfo{volume}{52}}, \bibinfo{pages}{263--379}
  (\bibinfo{year}{1979}).

\bibitem{fermi65}
\bibinfo{author}{Fermi, E.}, \bibinfo{author}{Pasta, J.} \&
  \bibinfo{author}{Ulam, S.}
\newblock \emph{\bibinfo{title}{{in Collected papers of E. Fermi, edited by E.
  Serge}}} (\bibinfo{publisher}{University of Chicago, Chicago},
  \bibinfo{year}{1965}).

\bibitem{berman05fermi}
\bibinfo{author}{Berman, G.} \& \bibinfo{author}{Izrailev, F.}
\newblock \bibinfo{title}{The fermi--pasta--ulam problem: fifty years of
  progress}.
\newblock \emph{\bibinfo{journal}{Chaos: An Interdisciplinary Journal of
  Nonlinear Science}} \textbf{\bibinfo{volume}{15}}, \bibinfo{pages}{015104}
  (\bibinfo{year}{2005}).

\bibitem{pettini90relaxation}
\bibinfo{author}{Pettini, M.} \& \bibinfo{author}{Landolfi, M.}
\newblock \bibinfo{title}{Relaxation properties and ergodicity breaking in
  nonlinear hamiltonian dynamics}.
\newblock \emph{\bibinfo{journal}{Physical Review A}}
  \textbf{\bibinfo{volume}{41}}, \bibinfo{pages}{768} (\bibinfo{year}{1990}).

\bibitem{bertini15prethermalization}
\bibinfo{author}{Bertini, B.}, \bibinfo{author}{Essler, F.~H.},
  \bibinfo{author}{Groha, S.} \& \bibinfo{author}{Robinson, N.~J.}
\newblock \bibinfo{title}{Prethermalization and thermalization in models with
  weak integrability breaking}.
\newblock \emph{\bibinfo{journal}{Physical Review Letters}}
  \textbf{\bibinfo{volume}{115}}, \bibinfo{pages}{180601}
  (\bibinfo{year}{2015}).

\end{thebibliography}

\end{document}